\RequirePackage{fix-cm}
\documentclass[twocolumn]{svjour3}          
\smartqed  
\usepackage{graphicx}
\usepackage{hyperref}
\usepackage{booktabs}
\usepackage{amsmath}
\usepackage{cuted}
\usepackage{amssymb}
\usepackage{float}
\usepackage{subfigure}
\usepackage{enumitem}
\usepackage[numbers,sort&compress]{natbib}
\usepackage{appendix}
\usepackage{enumitem}
\setlist[itemize]{label=\textbullet}
\makeatletter
\def\rubric{}
\makeatother
\renewcommand{\makeheadbox}{\relax}

\begin{document}

\title{Gravitational lensing and shadow around a non-minimally coupled Horndeski black hole in plasma medium
}

\titlerunning{Optical properties of non-minimally coupled Horndeski black hole in Plasma Medium}        

\author{Shubham Kala \and
       Jaswinder Singh 
}


\institute{Shubham Kala \at
              The Institute of Mathematical Sciences\\
              C.I.T. Campus, Taramani, Chennai-600113\\
              Tamil Nadu, India\\
              \email{shubhamkala871@gmail.com}
           \and
          Jaswinder Singh \at
          Department of Physics, Institute of Science\\
           Banaras Hindu University, Varanasi-221005, India\\
            \email{jaswinderbhu08@gmail.com}
            }
           
\date{Received: date / Accepted: date}
\maketitle
\begin{abstract}
We investigate the light deflection and the shadow characteristics of a non-minimally coupled Horndeski black hole surrounded by a magnetized, cold, pressureless plasma medium, while considering both homogeneous and non-homogeneous plasma distributions. We consider an analytical expression for the deflection angle of light and analyze how it is influenced by the plasma properties and the Horndeski coupling constant. The circular light orbits, which define the photon sphere, are also analyzed for both types of plasma media, highlighting their impact on the shadow boundary. The shadow properties of the black hole are examined in detail, and constraints on the model parameters are derived by comparing the theoretical shadow radius with observational measurements of Sgr A* and M87* obtained by the Event Horizon Telescope Collaboration. We also study the black hole shadow images along with the corresponding intensity profiles produced by a radially infalling accretion flow in the plasma environment. The results are particularly interesting, as they reveal how the modified black hole geometry affects both the plasma distribution and the black hole parameters in a realistic astrophysical context.
\end{abstract}

%
%
\maketitle
%
%
\section{Introduction}\label{sec:1}
Einstein’s geometric formulation of gravity, known as General Relativity (GR), has been subject to rigorous experimental scrutiny, and remains the cornerstone of modern gravitational theory \cite{Einstein:1916vd}. One of the most intriguing predictions of GR in four-dimensional $(4D)$ spacetime is the existence of black holes (BH), which are exact solutions to Einstein's field equations (EFE) \cite{Schwarzschild:1916uq}. Nevertheless, despite its empirical successes, GR encounters several conceptual and observational challenges, including the explanation of dark energy \cite{Planck:2013pxb}, the presence of singularities \cite{Penrose:1964wq}, and possible deviations in the strong--field regime \cite{Pang:2018hjb}. These issues have spurred the development of alternative theories of gravity that extend or modify the Einsteinian paradigm, among which are \( f(R) \) theories \cite{Sotiriou:2008rp}, \( f(R, T) \) theories \cite{Harko:2011kv}, generalized scalar--tensor theories \cite{fujii2003scalar}, and Tensor--Vector--Scalar theories \cite{moffat2006scalar}. In particular, Horndeski gravity, which represents the most general class of scalar--tensor theory with second order field equations, has received significant attention \cite{Horndeski:1974wa}. Horndeski gravity introduces an additional scalar degree of freedom coupled non-minimally to curvature invariants, thereby enriching the theoretical landscape while preserving freedom from Ostrogradsky instabilities \cite{Ostrogradsky:1850fid,Grosse-Knetter:1993tae,Horndeski:2024sjk}. In this context, studying the propagation of massless particles in the strong--field region around BHs offers a powerful means to probe the scalar field effects and spacetime modifications predicted by Horndeski gravity, enabling observational tests beyond GR.\\

One of the most striking predictions of GR is the bending of light in curved spacetime, a phenomenon known as gravitational lensing (GL) \cite{Dyson:1920cwa,Muhleman:1970zz}. Over the past few decades, GL has emerged as a powerful astrophysical and cosmological tool for probing the distribution of matter, both luminous and dark, across the Universe \cite{Paczynski:1985jf,Virbhadra:1998dy,Virbhadra:1999nm,Bartelmann:1999yn,Bozza:2001xd,Treu:2010uj,Nascimento:2020ime}. In the classical vacuum scenario, the deflection angle of light depends solely on the mass distribution of the lensing object and is independent of the photon's frequency. 
However, realistic astrophysical environments often include ionized plasma, whose dispersive properties introduce a frequency dependence on photon trajectories \cite{Neufeld:1955zz}. In such media, photons deviate from strictly null geodesics, and their propagation is effectively described by a refractive index that varies with position and frequency. Ehlers et al. \cite{Ehlers:1987nm} developed a Hamiltonian framework for light rays in a magnetized, pressureless plasma on curved backgrounds, while Tsupko et al. later simplified this derivation to a non-magnetized, pressureless plasma \cite{tsupko2012gravitational}. In this framework, the scalar refractive index encapsulates the influence of plasma and enables the derivation of photon motion equations, first investigated systematically by Synge \cite{Synge:1966okc}.\\

These methods have led to significant advances in understanding the influence of plasma on light deflection near BHs. Perlick et al. \cite{Perlick:2017fio} computed the deflection angle in Schwarzschild and Kerr spacetimes for plasmas with radially varying density. Similar investigations by Bisnovatyi-Kogan and Tsupko \cite{Tsupko:2009zz,Bisnovatyi-Kogan:2010flt,Tsupko:2013cqa} provided detailed analyses of plasma lensing effects. Morozova et al. \cite{morozova2013346} extended this to slowly rotating Kerr BHs on the equatorial plane. Subsequent studies explored the frequency-dependent deflection of light in various scenarios, including the works by Er and Mao \cite{Er:2013efa} and Rogers \cite{rogers2015frequency}. Building on these foundational works, recent research has focused on the impact of plasma on gravitational lensing around a wider variety of BH spacetimes, including dyonic ModMax BHs \cite{Turakhonov:2024smp}, regular and quantum-corrected BHs \cite{Alloqulov:2024xak}, dual-charged stringy BHs \cite{Kala:2024fvg}, and BHs in alternative theories of gravity. Recent investigations have revealed that plasma effects, together with BH parameters such as charge and coupling constants, systematically modify deflection angles and lensing observables in both strong- and weak-field regimes. Incorporating realistic plasma distributions and comparing with observational data have refined our theoretical models of gravitational lensing near compact objects. \cite{Bisnovatyi-Kogan:2015dxa,Jin:2020emq,Atamurotov:2021cgh,Atamurotov:2021imh,Crisnejo:2019ril,Babar:2021nst,Perlick:2023znh,Davlataliev:2023ckw,Kumar:2023wfp,Orzuev:2023dus,Atamurotov:2023rye,Atamurotov:2023tff,Li:2023esz,Ditta:2023lny,Molla:2022izk,Javed:2022fsn,Kala:2022uog,Vishvakarma:2024icz,Roy:2025hdw}.\\

The field of observational BH physics achieved a transformative breakthrough in 2019, when the Event Horizon Telescope (EHT) collaboration released the first horizon-scale image of the supermassive BH shadow in galaxy M87*~\cite{Akiyama:2019cqa,Akiyama:2019eap}. This striking observation revealed a bright emission ring surrounding a central dark region, the BH shadow, offering an unprecedented opportunity to test theoretical models of strong-field gravity. The foundational theoretical framework for BH shadows traces back to Synge \cite{Synge:1966okc}, who demonstrated that a spherically symmetric BH like Schwarzschild produces a perfectly circular shadow. Bardeen later showed that spin distorts this shadow into a D-shaped silhouette in rotating Kerr BHs \cite{bardeen1973black}. Building on these foundations, Perlick et al. \cite{Perlick:2015vta} analytically derived the angular size of BH shadows in spherically symmetric spacetimes surrounded by non-magnetized plasma. Atamurotov et al. \cite{atamurotov2015optical} extended these studies to rotating Kerr BHs embedded in plasma with radial power-law density, exploring the combined effects of spin and plasma. Subsequent work investigated whether plasma leaves detectable imprints on shadow shape and size, with significant efforts by several groups \cite{Huang:2018rfn,Jha:2021eww,Badia:2022phg,Briozzo:2022mgg,Atamurotov:2022iwj,Raza:2024zkp,Pantig:2024lpg}. These ongoing studies bridge theoretical insights with observations, and help elucidate the rich interplay between plasma physics and BH shadows within GR and beyond.\\

In this paper, we consider the Horndeski BH solution and study its optical properties in plasma medium. This class of BH solutions arises from scalar–tensor theories of gravity that extend GR by incorporating non--minimal couplings between a scalar field and curvature invariants \cite{Babichev:2017guv,babichev2024exact}. The motivation for selecting the Horndeski BH solution lies in its relevance to probe potential deviations from GR in strong gravitational fields. Furthermore, the rich parameter space of the Horndeski BH solution enables a detailed investigation of how scalar--field interactions influence observable phenomena such as photon sphere, light deflection and shadows in plasma environment \cite{Badia:2017art,Gao:2023mjb}. These features are particularly relevant in the context of ongoing and future next-generation astronomical observations, including those conducted by the EHT.\\

The paper is organized as follows. In section~\ref{sec:2}, we introduce the non-minimally coupled Horndeski BH and its geometry. In section~\ref{sec:3}, we formulate the governing equations for the propagation of the light in plasma medium. In sections~\ref{sec:4} and~\ref{sec:5}, we consider analytical expressions for light deflection in homogeneous and non-homogeneous plasma media, respectively. In section~\ref{sec:6} we analyze the circular photon orbits defining the BH shadow. In section~\ref{sec:7} we investigate how plasma parameters and coupling constants affect the shadow, while thes section~\ref{sec:8} presents simulated shadow images with an infalling gas distribution. Finally, in section~\ref{sec:9} we summarize the key results and discuss future research directions.
\section{Spacetime geometry of non-minimally coupled Horndeski black hole}
\label{sec:2}
The non-minimally coupled Horndeski BH belongs to a family of solutions in Horndeski gravity, which is the most general scalar–tensor theory leading to second-order field equations. In modern notation, the Horndeski action is written as~\cite{Horndeski:1974wa}:
\begin{align}
S = \int d^4 x \sqrt{-g} \bigl[
    & G_2(X)
    - G_3(X) \, \Box \varphi
    + G_4(X) R 
\nonumber \\
    & + G_{4,X}(X) 
      \bigl( (\Box \varphi)^2 - \varphi_{\mu\nu} \varphi^{\mu\nu} \bigr)
\nonumber \\
    & + G_5(X) G^{\mu\nu} \varphi_{\mu\nu}
\nonumber \\
    & - \tfrac{1}{6} G_{5,X}(X) 
      \bigl( (\Box \varphi)^3 
           - 3 \Box \varphi \, \varphi_{\mu\nu} \varphi^{\mu\nu} 
\nonumber \\
    & \qquad {} + 2 \varphi_{\mu\nu} \varphi^{\nu\rho} \varphi_\rho^{\;\mu} 
      \bigr)
\bigr].
\end{align}
Here, \( G_{i,X} \equiv \frac{\partial G_i}{\partial X} \) and \( \varphi_{\mu\nu} \equiv \nabla_\mu \nabla_\nu \varphi \). 

\noindent By specifying the Horndeski functions in the original Lagrangian as
\[
G_2 = \eta X, 
\quad G_4 = 1 + \beta \sqrt{-X}, 
\quad G_3 = G_5 = 0,
\]
we obtain an effective action given as,
\begin{align}
S = \int d^4 x \sqrt{-g} \biggl[
    &\Bigl( 1 + \beta \sqrt{ \frac{(\partial \varphi)^2}{2} } \Bigr) R
    - \frac{\eta}{2} (\partial \varphi)^2 
    \notag \\
    &\quad - \beta \sqrt{ 2 (\partial \varphi)^2 }
    \Bigl( (\Box \varphi)^2 - (\nabla_\mu \nabla_\nu \varphi)^2 \Bigr)
\biggr],
\end{align}
where where the $G_{2}$ term is a canonical kinetic term, and the constant 1 in $G_{4}$ corresponds to a pure Einstein-Hilbert term. \( \phi \) is the scalar field , \( R \) denotes the Ricci scalar, and \( \eta \) and \( \beta \) are coupling constants of the theory. 

\begin{figure}[htbp]
	\begin{center}
		{\includegraphics[width=0.45\textwidth]{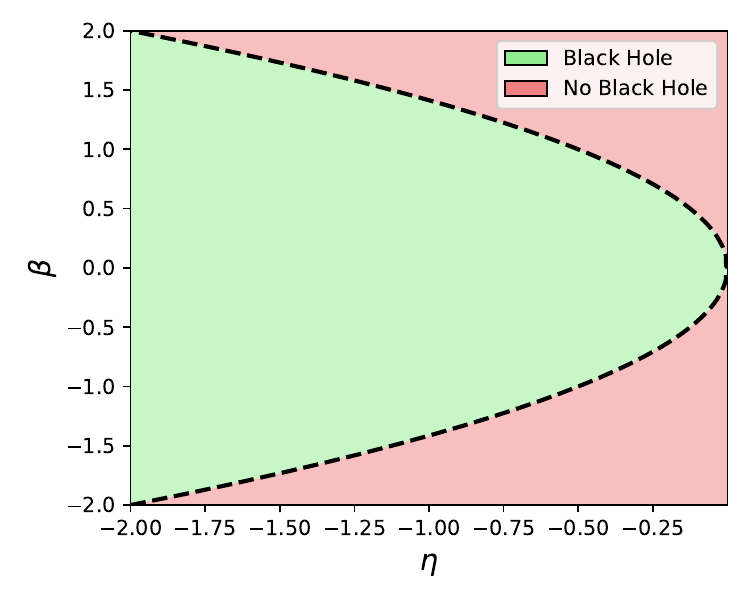}}
         \end{center}
	\caption{Parameter space $(\beta, \eta)$ representing a quartic squre-root Horndeski BH. Black dashed line corresponds to extreme BHs. The light green region represents the range of parameter values where BHs exist.} \label{fig:bhANDnobh}
\end{figure}

The theory admits static, spherically symmetric asy-mptotically flat BH solution in $4D$ spacetime known as quartic square-root Horndeski BH given by the line element \cite{Babichev:2017guv},  
\begin{equation}
ds^2 = -f(r)\, dt^2 + \frac{dr^2}{f(r)} + r^2 d\Omega_2^2.
\end{equation}
The function $f(r)$  is given by,   
\begin{equation}
f(r) = 1 - \frac{2M}{r} - \frac{\beta^2}{2 \eta r^2},
\end{equation}
where \( M \) is an integration constant interpreted as the BH mass. Here, $\beta$ and $\eta$ are coupling parameters arising from the scalar field sector of the Horndeski theory, where $\beta$ effectively plays the role of a scalar charge, and $\eta$ characterizes the strength and sign of the non-minimal coupling between the scalar field and gravity. 
Fig.~\ref{fig:bhANDnobh} shows the allowed ranges of $\beta$ and $\eta$, illustrating the parameter space that separates BH solutions from non–BH configurations. The solution is physically consistent only when $\eta$ and $\beta$ share the same sign. Therefore, to ensure consistency of the solution throughout the manuscript, we choose $\beta < 0$.

The event horizon(s) of this BH are determined by the real, positive roots of the equation \( f(r) = 0 \). 
Solving for \( r \), one finds:
\begin{equation}
r_H = M \pm \sqrt{ M^2 + \frac{\beta^2}{2\eta} }.
\end{equation}
This expression shows that the existence and position of the horizon depend on both the scalar coupling constant \( \beta \) and the parameter \( \eta \).
In the limit $\beta \to 0$, the horizon reduces to $r=2M$, which corresponds to the Schwarzschild geometry. At spatial infinity, the spacetime remains asymptotically flat, while a curvature singularity persists at $r=0$. For $\eta>0$, the geometry admits a single horizon. Conversely, for $\eta<0$, the existence of horizons depends on the magnitude of $\beta$: 
\begin{itemize}
    \item If $M < \frac{|\beta|}{\sqrt{-2\eta}}$, there is no horizon, resulting in a naked singularity.
    \item If $M > \frac{|\beta|}{\sqrt{-2\eta}}$, the solution exhibits two distinct horizons.
\end{itemize}
\begin{figure}[htbp]
	\begin{center}
		{\includegraphics[width=0.45\textwidth]{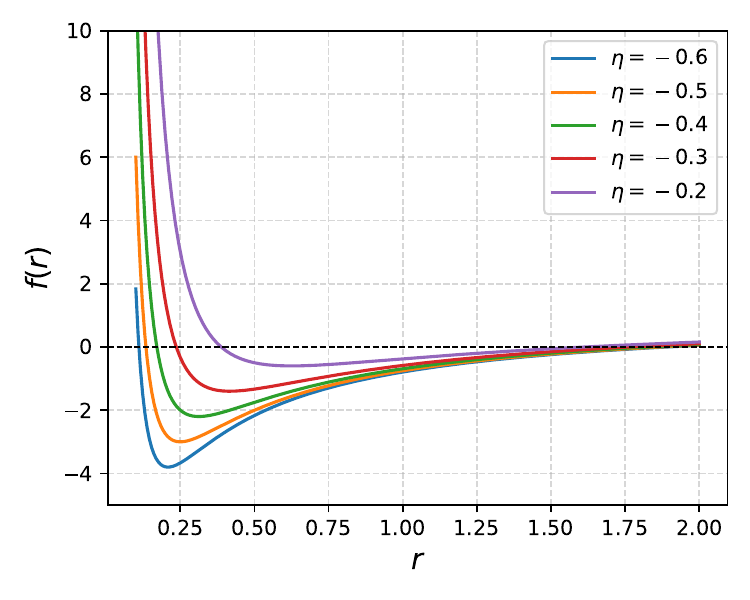}}
            {\includegraphics[width=0.45\textwidth]{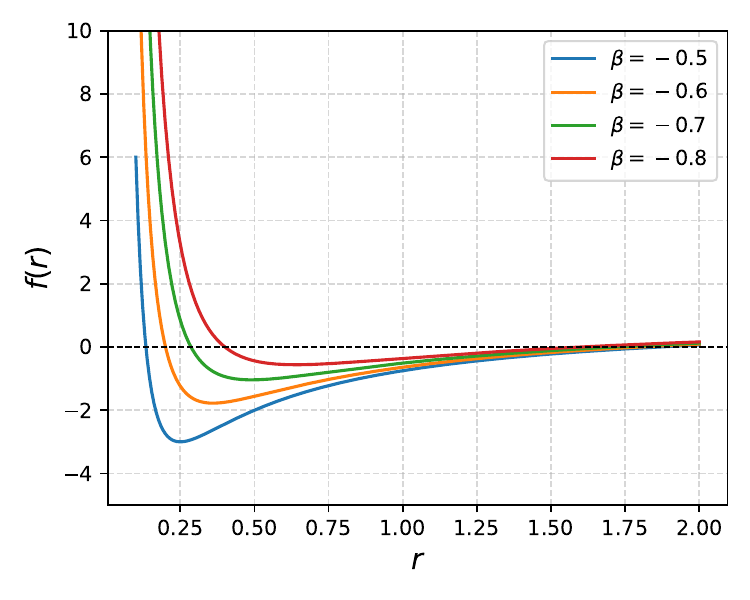}}
	\end{center}
	\caption{The horizons of BHs for different values of $\eta$ (upper panel) with $M=1$ and $\beta=-0.5$ and $\beta$ (lower panel) $M=1$ and $\eta=-0.5$.} \label{fig:frvsr}
\end{figure}
Fig.~\ref{fig:frvsr} illustrates the variation of the metric function $f(r)$ with respect to the radial coordinate $r$. Throughout, we impose the condition $M > \frac{|\beta|}{\sqrt{-2\eta}}$, which, as evident in the plots, leads to the existence of two distinct horizons.
\section{Light propagation in plasma medium}\label{sec:3}
\noindent
Consider a general static, spherically symmetric spacetime given by
\begin{equation}
ds^{2} = -A(r) dt^{2} + B(r) dr^{2} + D(r) \bigl( d\theta^{2} + \sin^{2}\theta d\varphi^{2} \bigr),
\end{equation}
with $A(r), B(r)>0$ outside the horizon. We analyze photon propagation in a non-magnetized, pressureless plasma characterized by plasma frequency 
\[
\omega_p^2(r) = \frac{4\pi e^2}{m_e} N(r).
\]
Here, $e$ is the charge of the electron, $m_{e}$ is the mass of the electron, and $N(r)$ is the number density of the electrons. 

\noindent The refractive index for a photon of conserved energy $\omega_0=-p_t$ (measured at spatial infinity) becomes
\begin{equation}
n^2(r,\omega)=1-\frac{\omega_p^2(r)}{\omega^2(r)},
\end{equation}
where $\omega(r)=\omega_0/\sqrt{A(r)}$ follows from the gravitational redshift. Using spherical symmetry, we restrict ourselves to the equatorial plane ($\theta=\pi/2$). The Hamiltonian governing photon motion in plasma then reads
\begin{equation}
H=\frac12\bigl[ g^{\mu\nu}p_\mu p_\nu + \omega_p^2(r)\bigr]=0.
\end{equation}
Explicitly, this becomes
\begin{equation}
-\frac{\omega_0^2}{A(r)} + \frac{p_r^2}{B(r)} + \frac{L^2}{D(r)} + \omega_p^2(r) = 0,
\end{equation}
where $L=p_\varphi$ is the conserved angular momentum.

\noindent From this, the condition for photon propagation at radius $r$ becomes
\begin{equation}
\omega^2(r) > \omega_p^2(r),
\end{equation}
meaning the photon frequency measured locally must exceed the local plasma frequency.

\noindent For the trajectory, using $\dot{r}/\dot{\varphi}$ yields
\begin{equation}
\frac{dr}{d\varphi} = \frac{D(r)}{B(r)} \frac{p_r}{L}.
\end{equation}
Eliminating $p_r$ using the Hamiltonian constraint, we obtain the orbit equation:
\begin{equation}
\frac{dr}{d\varphi} = \pm \frac{\sqrt{D(r)}}{\sqrt{B(r)}} 
\sqrt{\frac{\omega_0^2}{L^2}h^2(r)-1},
\end{equation}
with the effective function
\begin{equation}
h^2(r)=\frac{D(r)}{A(r)} \left[1 - A(r)\,\left(\frac{\omega_p^2(r)}{\omega_0^2}\right)\right].
\end{equation}
Here, $h(r)$ encodes the combined effects of geometry and plasma. 

\section[Light deflection with constant plasma frequency]{Light deflection by homogeneous plasma medium 
}\label{sec:4}
\noindent
In this section, we study gravitational lensing of photons propagating in a homogeneous plasma with constant plasma frequency $\omega_p$, i.e., $\omega_p^2(r)=\omega_p^2=\mathrm{const}$. The refractive index then becomes
\[
n^2(r)=1-\frac{\omega_p^2}{\omega^2(r)},
\]
where $\omega(r)=\omega_0/\sqrt{A(r)}$.

\noindent A light ray coming from infinity, reaching a closest approach distance $R$, and then returning to infinity, will accumulate a total deflection angle $\delta$ given by \cite{Bisnovatyi-Kogan:2010flt},
\begin{equation}
\delta + \pi = 2 \int_{R}^{\infty} \frac{\sqrt{B(r)}}{\sqrt{D(r)}} 
\left[\frac{\omega_0^2}{L^2} h^2(r)-1\right]^{-1/2} dr,
\end{equation}
\begin{figure}[H]
	\begin{center}
		{\includegraphics[width=0.45\textwidth]{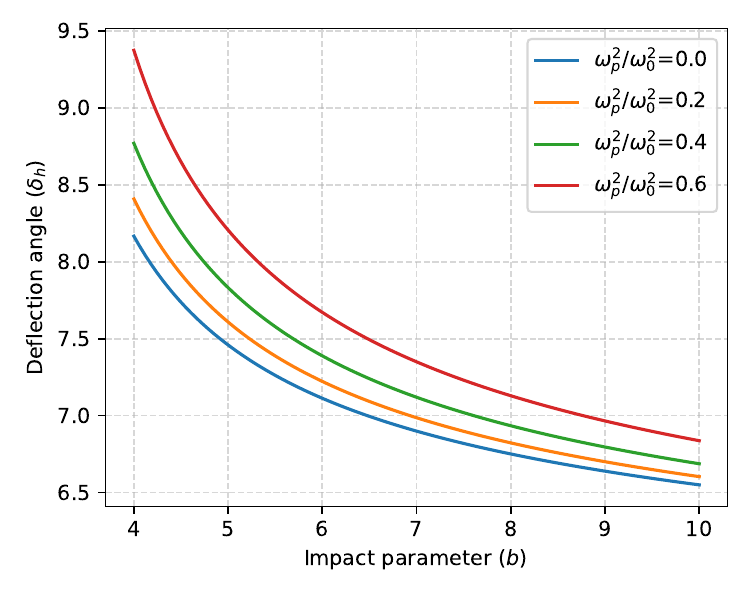}}
      {\includegraphics[width=0.45\textwidth]{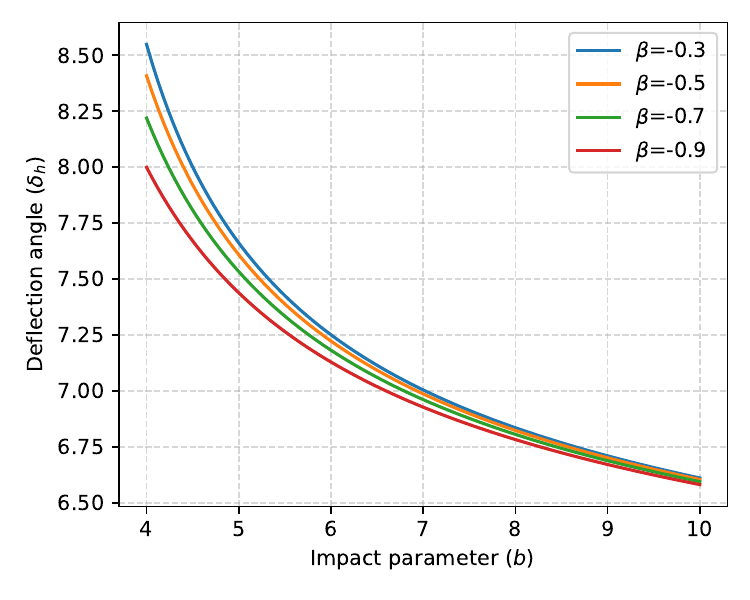}}
       {\includegraphics[width=0.45\textwidth]{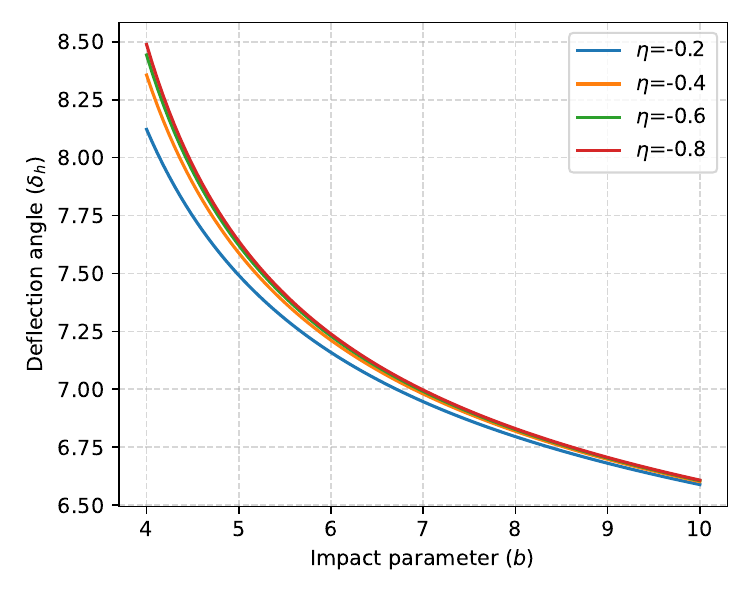}}
	\end{center}
	\caption{Variation of the deflection angle as a function of the impact parameter in a homogeneous plasma medium. The upper panel shows the effect of varying the homogeneous plasma parameter $\omega_p^2/\omega_0^2$ for fixed $\eta=-0.5$ and $\beta=-0.5$.The middle panel illustrates the variation with different values of $\beta$ for fixed $\eta=-0.5$ and $\omega_p^2/\omega_0^2=0.4$. The lower panel presents the variation with different values of $\eta$ for fixed $\beta=-0.5$ and $\omega_p^2/\omega_0^2=0.4$. } \label{fig:deflectionhomo}
\end{figure}
where the effective function is given by,
\[
h^2(r)=\frac{D(r)}{A(r)} \Bigl[1 - A(r)\,\omega_p^2/\omega_0^2\Bigr].
\]
At the turning point \( r = R \), the radial velocity vanishes, leading to \( h^2(R) = L^2/\omega_0^2 \) from the orbit equation.

\noindent This allows the bending angle to be expressed entirely in terms of $R$, as follows
\begin{equation} \label{BAExp}
\delta + \pi = 2 \int_{R}^{\infty} \frac{\sqrt{B(r)}}{\sqrt{D(r)}} 
\Bigl[\frac{h^2(r)}{h^2(R)}-1\Bigr]^{-1/2} dr.
\end{equation}
This is the expression for the bending angle of light in a general spherically symmetric spacetime in the presence of a plasma medium. We will use this integral to compute the deflection angle for both homogeneous and non-homogeneous plasma distributions. 
\begin{figure}[H]
	\begin{center}
		{\includegraphics[width=0.45\textwidth]{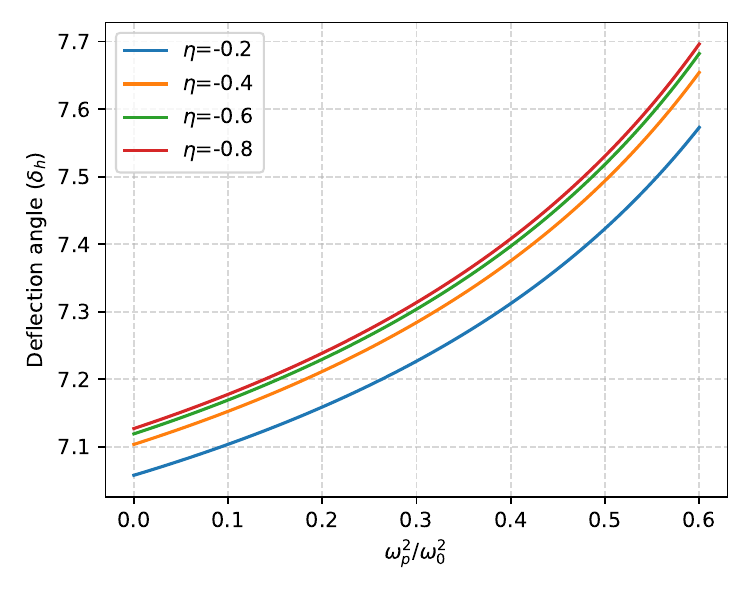}}
            {\includegraphics[width=0.45\textwidth]{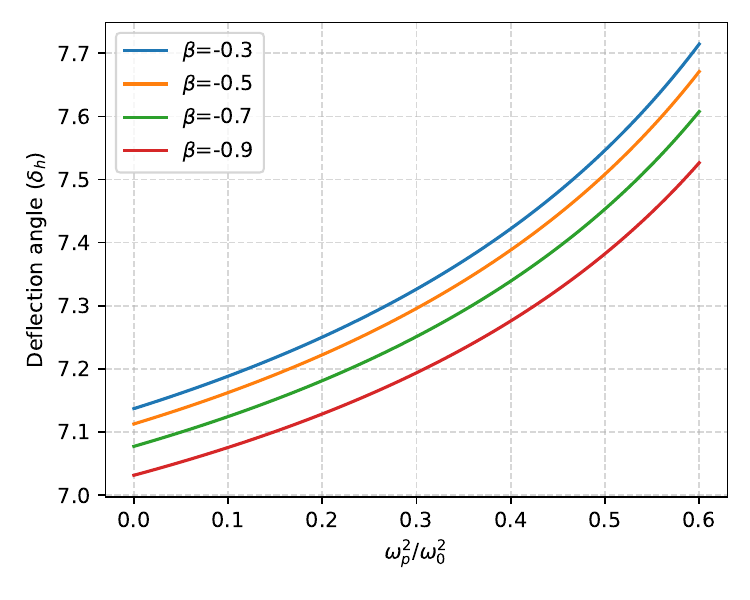}}
	\end{center}
	\caption{Variation of the deflection angle as a function of the homogeneous plasma parameter $(\omega_p^2/\omega_0^2)$. Upper panel: results for different values of $\eta$ with $M=1$ and fixed $\beta=-0.5$. Lower panel: results for different values of $\beta$ with $M=1$ and fixed $\eta=-0.5$.} \label{fig:deflectionhomo2}
\end{figure}
\noindent
For the specific case of the non-minimally coupled Horndeski BH, the metric functions are given by:
\[
A(r)=f(r)=1-\frac{2M}{r}-\frac{\beta^2}{2\eta r^2}, 
\quad B(r)=\frac{1}{f(r)}, 
\quad D(r)=r^2.
\]
Therefore, the corresponding function $h^2(r)$ can be expressed as
\[
h^2(r)=\frac{r^2}{f(r)}\bigl[1 - f(r)\,\omega_p^2/\omega_0^2\bigr].
\]
Substituting these into the integral in Eq.~\ref{BAExp} yields the explicit expression for the bending angle of light in a homogeneous plasma medium, given by
\begin{equation}
\delta + \pi = 2 \int_{R}^{\infty} \frac{dr}{r\sqrt{f(r)}} 
\left[\frac{r^2f(r)^{-1}\bigl(1 - f(r)\,\omega_p^2/\omega_0^2\bigr)}{h^2(R)}-1\right]^{-1/2}.
\end{equation}
where $R$ is the turning point of the trajectory determined by $h(R)^2 = b^2$, and $b$ is commonly known as the impact parameter. We have solved this expression numerically, and the results are discussed in the subsequent section.

\noindent Fig.~\ref{fig:deflectionhomo} illustrates the variation of the deflection angle as a function of the impact parameter in a homogeneous plasma medium. The upper panel shows the effect of increasing the homogeneous plasma parameter $\omega_p^2/\omega_0^2$ for fixed $\eta=-0.5$ and $\beta=0.5$. It is evident that the deflection angle increases with higher plasma density (shown more clearly in Fig.~\ref{fig:deflectionhomo2}), since the plasma raises the effective refractive index around the BH, leading to stronger bending of light rays. The middle panel presents the variation for different values of $\beta$ with fixed $\eta=-0.5$ and $\omega_p^2/\omega_0^2=0.4$, showing that the deflection angle is smaller for lower $\beta$. A higher $\beta$ effectively contributes more to the gravitational potential, thus enhancing the bending. The lower panel shows the variation for different values of $\eta$ with fixed $\beta=0.5$ and $\omega_p^2/\omega_0^2=0.4$, where the deflection angle increases as $\eta$ becomes more negative. A more negative $\eta$ amplifies the gravitational potential, producing greater curvature of light paths. Overall, as expected from general relativistic lensing, the deflection angle decreases with increasing impact parameter.
\section[Light deflection in non-homogeneous plasma]%
{Light deflection by non-homogeneous plasma medium 
}\label{sec:5}
In the following, we extend our analysis to the case of a non-homogeneous plasma medium, where the plasma frequency varies with the radial coordinate according to a power-law profile, $\omega_{p}(r)^{2} = P_{0}/r^{k}$, with $P_{0}$ and $k \ge 0$ being arbitrary constants~\cite{rogers2015frequency,er2018two}.
\begin{figure}[H]
	\begin{center}
		{\includegraphics[width=0.45\textwidth]{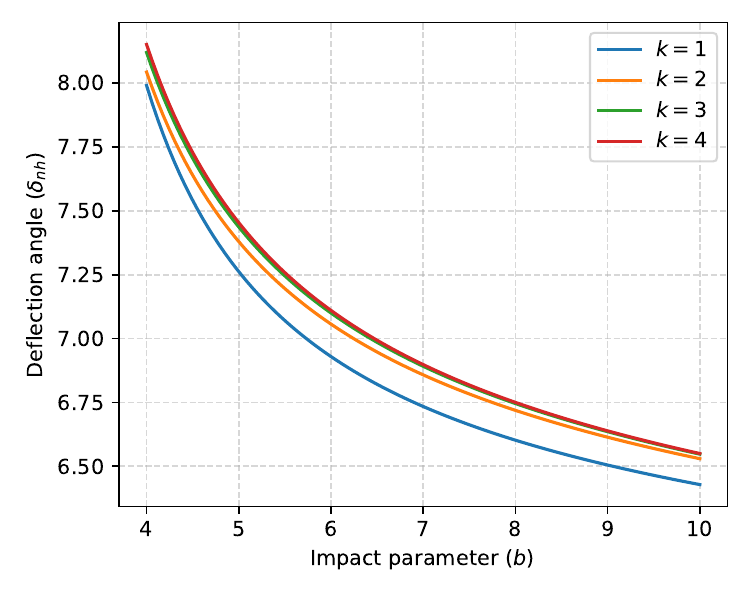}}
            {\includegraphics[width=0.45\textwidth]{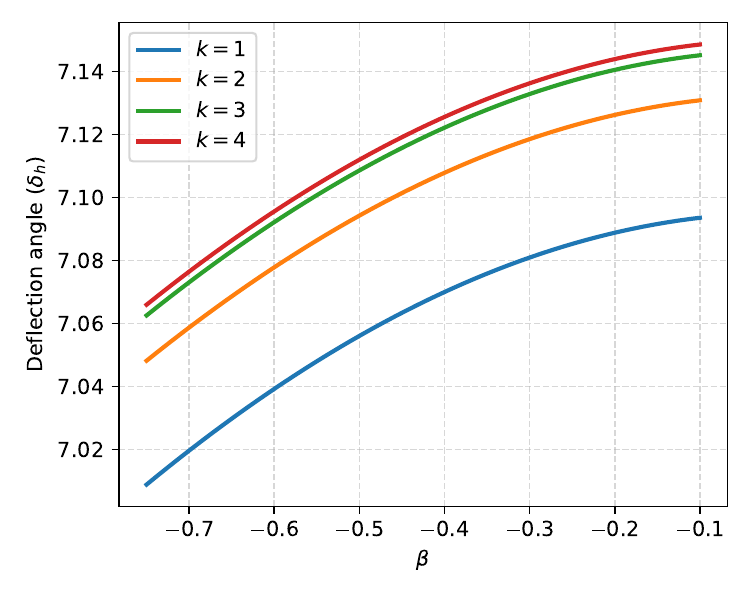}}
            {\includegraphics[width=0.45\textwidth]{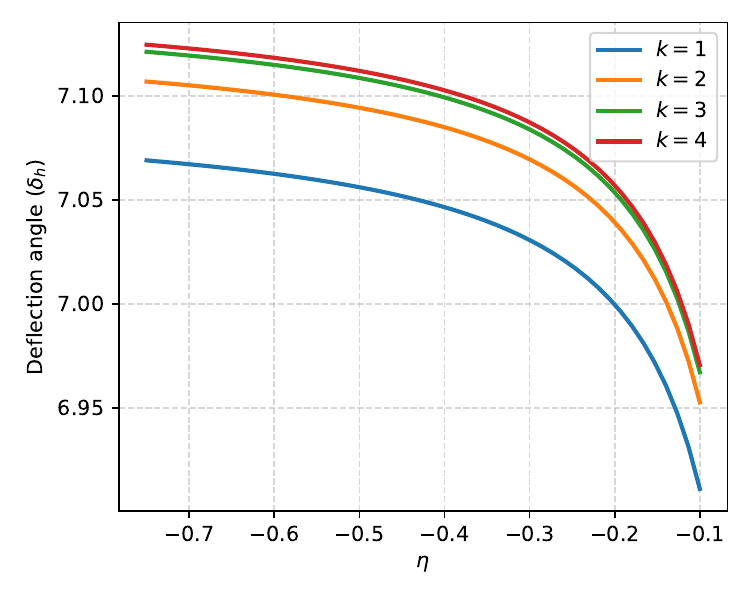}}
	\end{center}
	\caption{Variation of deflection angle as a function of $b$, $\beta$ and $\eta$. Here we consider $M=1$ and $P_0/\omega_0^2=0.5$. (Upper panel) Different values of $k$ with $\eta=-0.5$, $\beta=-0.5$; (Middle panel) Different values of $k$ with  $\eta=-0.5$; (Lower panel) Different values of $k$ with $\beta=-0.5$.} \label{fig:deflectionnonhomo}
\end{figure}
This form models realistic astrophysical scenarios where the plasma density decreases with distance from the BH. Incorporating this into the photon propagation modifies the effective optical geometry experienced by light rays. Specifically, the function $h(r)$, which governs the trajectory, becomes
\begin{equation}
    h(r)^2 = \frac{D(r)}{A(r)} \left[1 - \frac{A(r) P_{0}}{\omega_{0}^{2} r^{k}}\right],
\end{equation}
making the bending angle sensitive to both the strength ($P_{0}$) and the radial fall-off rate ($k$) of the plasma. 

The deflection angle in this non-homogeneous medium retains the integral form,
\begin{equation}
    \delta + \pi = 2 \int_{R}^{\infty} \frac{\sqrt{B(r)}}{\sqrt{D(r)}}
    \left[ \frac{h(r)^{2}}{h(R)^{2}} - 1 \right]^{-\frac12} dr.
\end{equation}
Similar to the homogeneous plasma case, we have also solved this expression numerically. This formulation provides a systematic way to explore how different plasma distributions influence the bending of light near compact objects.\\
Figure~\ref{fig:deflectionnonhomo} illustrates the variation of the deflection angle as a function of the impact parameter $b$, the parameter $\beta$, and $\eta$, considering $M=1$ and $P_0/\omega_0^2=0.5$. The upper panel shows the effect of different values of the non-homogeneous plasma parameter $k$ for fixed $\eta=-0.5$ and $\beta=-0.5$. The middle panel presents the variation of the deflection angle with respect to $\beta$ for different values of $k$ at fixed $\eta=-0.5$. The lower panel shows the variation with respect to $\eta$ for different $k$ at fixed $\beta=-0.5$. The deflection angle decreases with increasing impact parameter, similar to the homogeneous case. A higher value of the non-homogeneous plasma parameter $k$ produces a larger deflection angle due to an enhanced effective refractive index around the BH, leading to stronger bending of light. When the deflection angle is plotted as a function of $\beta$, it increases with larger $\beta$, while it decreases with increasing $\eta$. Nevertheless, the overall deflection remains maximal for higher values of $k$, highlighting the significant influence of plasma distribution on gravitational lensing.

\section{Circular light orbits in plasma medium}\label{sec:6}
The determination of the photon sphere, where massless particles follow circular trajectories, is essential for describing the BH shadow. Originally discussed by Atkinson~\cite{atkinson1965light}, the photon sphere corresponds to the set of unstable circular light orbits that act as critical boundaries for incoming rays. For asymptotically flat spacetimes where $\omega_p(r) \rightarrow 0$ as $r \rightarrow \infty$, the outermost photon sphere is typically unstable under radial perturbations, meaning that light rays can asymptotically spiral towards it but cannot remain stably confined. If a light ray coming from infinity reaches a closest distance $R > r_p$, it will return to infinity, whereas the critical case $R = r_p$ corresponds to the light ray asymptotically approaching the photon sphere of radius $r_p$. In scenarios where additional photon spheres exist, light rays can traverse and potentially return, but here we focus on the outermost photon sphere. Following the approach of Perlick et al.~\cite{Perlick:2015vta}, the condition defining the radius $r_p$ of a circular photon orbit is obtained by extremizing the optical impact function $h(r)$:
\begin{equation}
    \frac{d}{dr} h(r)^{2}\bigg|_{r=r_p} = 0.
\end{equation}
This equation can be solved separately for the homogeneous and non-homogeneous plasma distributions considered earlier. Each solution yields the modified photon sphere radius $r_p$ corresponding to the presence of plasma.

\begin{figure}[H]
	\begin{center}
		{\includegraphics[width=0.44\textwidth]{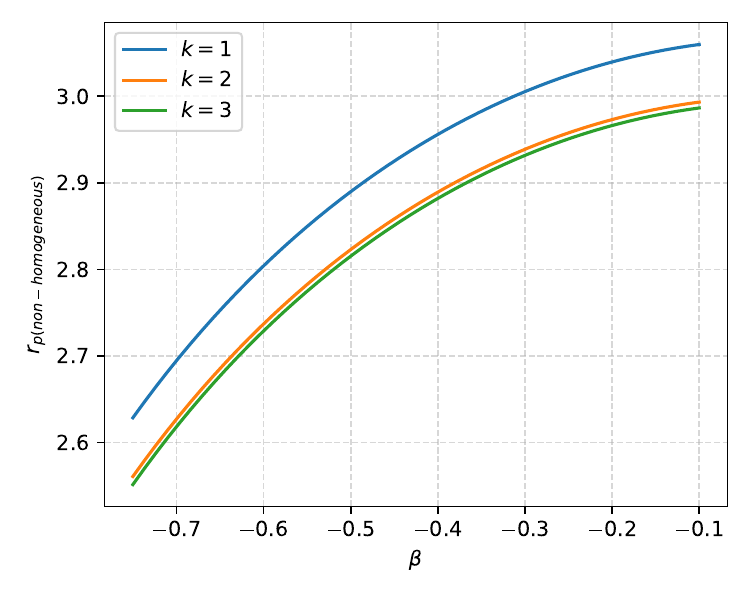}}
            {\includegraphics[width=0.44\textwidth]{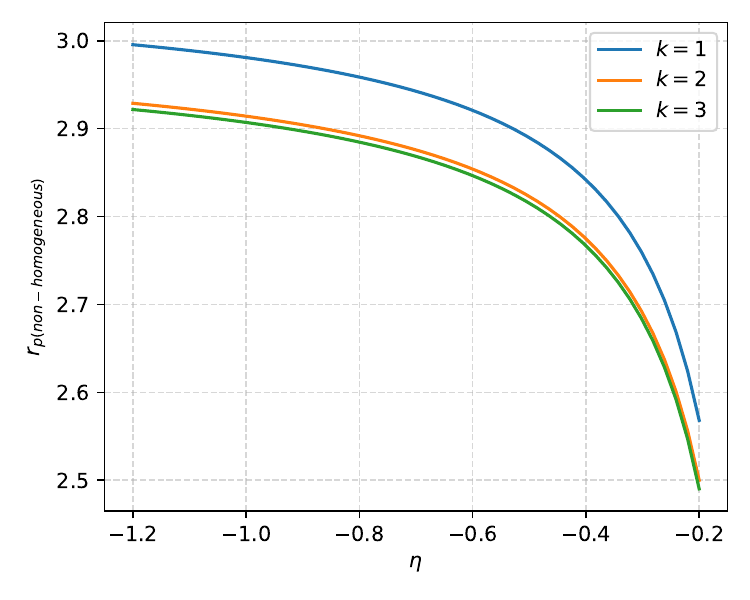}}
	\end{center}
	\caption{Variation of the photon sphere radius with the non-homogeneous plasma parameter. 
(Upper panel) As a function of $\beta$ with $M=1$ and $\eta=-0.5$; 
(Lower panel) As a function of $\eta$ with $M=1$ and $\beta=-0.5$.} \label{fig:rphnonhomogeneous}
\end{figure}

\noindent Explicitly, for homogeneous plasma medium
\[
h(r)^2 = \frac{D(r)}{A(r)} \Bigl[1 - A(r) \frac{\omega_p(r)^2}{\omega_0^2} \Bigr],
\]
differentiation w.r.t. r gives
\begin{multline}
\frac{d}{dr} h(r)^{2} = \frac{D'(r)}{A(r)} \Bigl[1 - A(r) \frac{\omega_p(r)^2}{\omega_0^2} \Bigr] \\
- \frac{D(r) A'(r)}{A(r)^2} \Bigl[1 - A(r) \frac{\omega_p(r)^2}{\omega_0^2} \Bigr] \\
- \frac{D(r)}{A(r)} \Bigl[ A'(r) \frac{\omega_p(r)^2}{\omega_0^2} + A(r) \frac{2 \omega_p(r) \omega_p'(r)}{\omega_0^2} \Bigr].
\end{multline}
Setting this expression equal to zero at $r=r_p$ yields the photon sphere radius in the presence of plasma,
which gives
\begin{multline}
\frac{2 r_{p}}{f(r_{p})} \biggl[ 1 - f(r_{p}) \frac{\omega_p(r_{p})^{2}}{\omega_{0}^{2}} \biggr]
- \frac{r_{p}^{2} f'(r_{p})}{f(r_{p})^{2}} \biggl[ 1 - f(r_{p}) \frac{\omega_p(r_{p})^{2}}{\omega_{0}^{2}} \biggr] \\
- \frac{r_{p}^{2}}{f(r_{p})} \biggl[ f'(r_{p}) \frac{\omega_p(r_{p})^{2}}{\omega_{0}^{2}} 
+ f(r_{p}) \frac{2 \omega_p(r_{p}) \omega_p'(r_{p})}{\omega_{0}^{2}} \biggr] = 0.
\end{multline}
The analytical solution of this equation is rather cumbersome; therefore, we solve it numerically to investigate the effect of the plasma medium on the photon sphere.
\begin{figure}[H]
	\begin{center}
		{\includegraphics[width=0.45\textwidth]{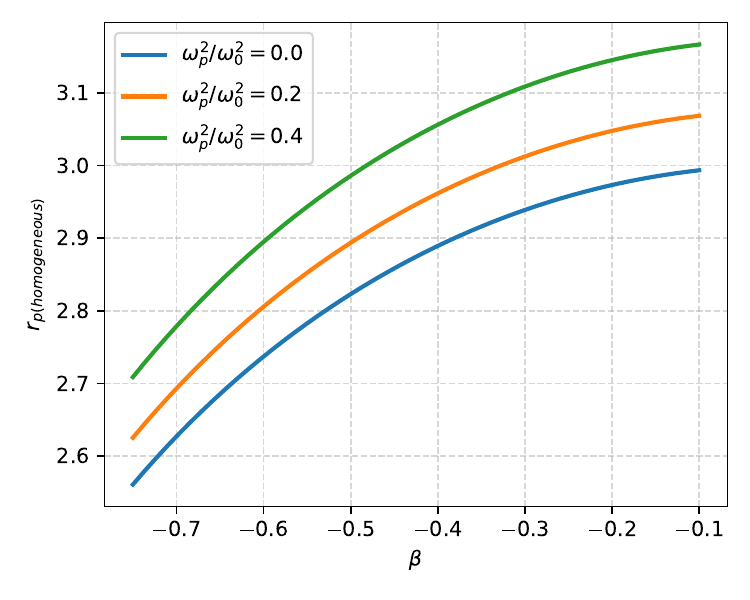}}
            {\includegraphics[width=0.45\textwidth]{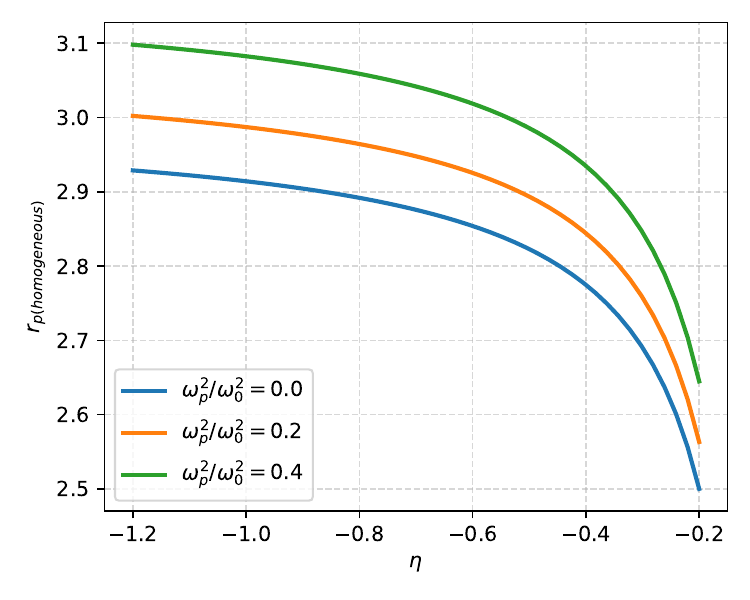}}
	\end{center}
	\caption{Variation of the photon sphere radius with the homogeneous plasma parameter. 
(Upper panel) As a function of $\beta$ with $M=1$ and $\eta=-0.5$; 
(Lower panel) As a function of $\eta$ with $M=1$ and $\beta=-0.5$.} \label{fig:rphhomogeneous}
\end{figure}
\noindent
Fig.~\ref{fig:rphhomogeneous} illustrates the variation of the photon sphere radius as a function of $\beta$ and $\eta$ for different values of the plasma parameter. It is observed that the photon sphere radius increases with increasing $\eta$ and decreases with increasing $\beta$. Physically, a higher value of $\eta$ effectively reduces the gravitational attraction near the BH, allowing photons to orbit farther out, thus increasing the photon sphere radius. Conversely, a larger magnitude of $\beta$ strengthens the gravitational pull, causing the photon sphere to shrink inward. Additionally, the photon sphere radius becomes larger for higher values of the homogeneous plasma parameter, since the increased plasma density raises the effective refractive index, leading to stronger bending of light and permitting circular photon orbits at larger radii. Similarly, the governing equation that determines the photon sphere in the presence of a non-homogeneous plasma medium is given by
\begin{multline}
\frac{2 r_{p}}{f(r_{p})} \biggl[ 1 - f(r_{p}) \frac{P_{0}}{\omega_{0}^{2} r_{p}^{k}} \biggr]
- \frac{r_{p}^{2} f'(r_{p})}{f(r_{p})^{2}} \biggl[ 1 - f(r_{p}) \frac{P_{0}}{\omega_{0}^{2} r_{p}^{k}} \biggr] \\
- \frac{r_{p}^{2}}{f(r_{p})} \cdot \frac{P_{0}}{\omega_{0}^{2}} 
\biggl[ \frac{f'(r_{p})}{r_{p}^{k}} - \frac{k f(r_{p})}{r_{p}^{k+1}} \biggr] = 0.
\end{multline}
Once again, due to the analytical complexity of the equation, we resort to numerical solutions, and the corresponding results are illustrated in Fig.~\ref{fig:rphnonhomogeneous}.
\vspace{-0.7cm}
\noindent The variation of the photon sphere radius shows a similar qualitative trend as in the homogeneous case when varying $\beta$ and $\eta$. However, it is notable that the photon sphere radius attains its minimum value for the highest values of the non-homogeneous plasma parameter $k$. 
This reflects the fact that in a non-homogeneous plasma distribution, increasing $k$ effectively concentrates the plasma closer to the BH, which enhances the local gravitational lensing effect and shifts the photon sphere inward.

\section{BH shadows surrounded by plasma medium}\label{sec:7}
In this section, we analyze the shadow cast by the non-minimally coupled Horndenski BH in the presence of a plasma medium. 
Following the standard geometric approach, the angular radius of the shadow as seen by a static observer at radial coordinate $r_{0}$ can be expressed as \cite{Synge:1966okc,Perlick:2015vta},
\begin{equation}
    \sin^{2} \alpha_{sh} = \frac{h^{2}(r_{p})}{h^{2}(r_{0})},
\end{equation}
where $r_{p}$ is the radius of the photon sphere determined by the circular light orbit condition. 
In the limit where the observer is located far from the BH ($r_{0} \to \infty$), the metric becomes asymptotically flat so that $A(r_{0}) \to 1$ and $D(r_{0}) \to r_{0}^{2}$. Furthermore, the plasma frequency vanishes at infinity, i.e., $\omega_p(r_{0}) \to 0$, leading to
\[
h(r_{0})^2 \approx r_{0}^{2}.
\]
\noindent Consequently, the angular radius simplifies to
\[
\sin \alpha_{sh} \approx \frac{h(r_{p})}{r_{0}}.
\]
For small angles, $\alpha_{sh} \approx \sin \alpha_{sh}$, so the shadow radius as measured by the distant observer becomes
\begin{equation}
    R_{sh} = r_{0} \alpha_{sh} \approx h(r_{p}).
\end{equation}
Specializing to our spherically symmetric metric with 
\[
A(r) = f(r), 
\quad D(r)=r^{2},
\]
we obtain
\begin{equation}
    R_{sh} = \sqrt{ \frac{r_{p}^{2}}{f(r_{p})} \bigl[ 1 - f(r_{p}) \frac{\omega_p(r_{p})^{2}}{\omega_0^2} \bigr] }.
\end{equation}
This expression gives the shadow radius in terms of the photon sphere radius $r_{p}$, the metric function $f(r)$ evaluated at $r_{p}$, and the local plasma frequency.
\begin{figure}[H]
	\begin{center}
		{\includegraphics[width=0.4\textwidth]{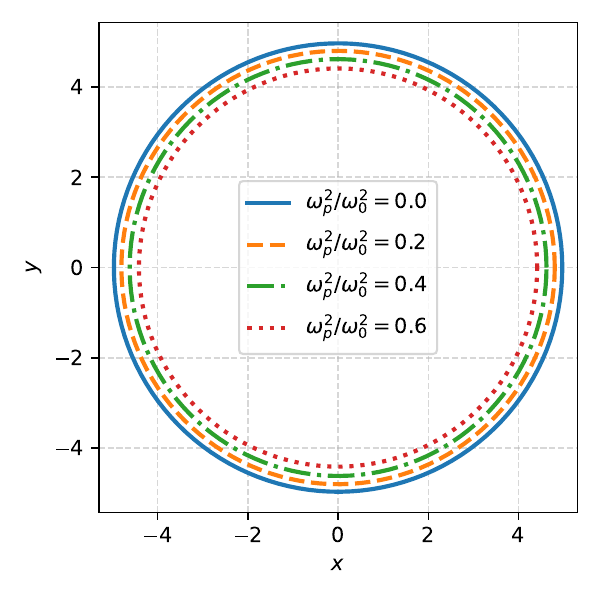}}
            {\includegraphics[width=0.4\textwidth]{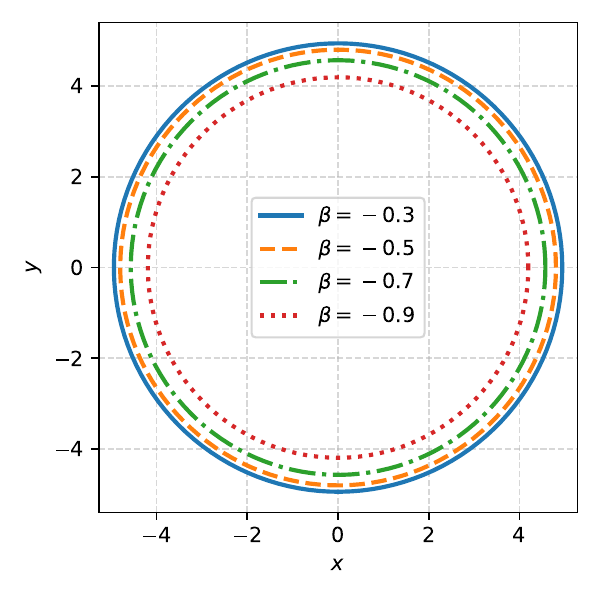}}
            {\includegraphics[width=0.4\textwidth]{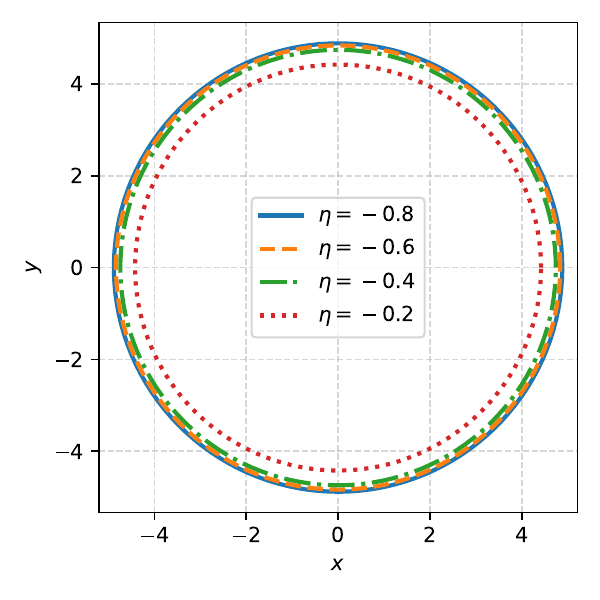}}
	\end{center}
	\caption{Variation of the shadow radius for different values of the homogeneous plasma parameter, $\beta$, and $\eta$ with $M=1$. 
We fix $\omega_p^2/\omega_0^2=0.5$, $\beta=-0.5$, and $\eta=-0.5$ where these parameters are not varied in the respective plots.
} \label{fig:shadowhomogeneous}
\end{figure}
In Fig.~\ref{fig:shadowhomogeneous}, we illustrate the variation of the shadow radius for the homogeneous plasma medium. 
Since the BH is non-rotating, the shadow remains perfectly circular; however, its size changes depending on the plasma distribution and the spacetime parameters in the BH solution. 
It is observed that increasing the homogeneous plasma parameter $\omega_p^2/\omega_0^2$ or the coupling parameter $\eta$ leads to a decrease in the shadow radius. 
This occurs because a denser plasma or higher $\eta$ effectively increases the refractive index around the BH, causing photons to bend more tightly and reducing the apparent shadow size. 
Conversely, incr easing the parameter $\beta$ enlarges the shadow radius, as a higher $\beta$ enhances the gravitational potential, allowing photons to orbit farther from the BH, thus increasing the apparent shadow size.

For the case of a non-homogeneous plasma medium described by a power-law plasma frequency profile,
\[
\omega_p(r)^{2} = \frac{P_{0}}{r^{k}},
\]
with $k \ge 0$, the function $h(r)$ becomes
\[
h(r)^{2} = \frac{r^{2}}{f(r)} \biggl[ 1 - f(r) \frac{P_{0}}{\omega_{0}^{2} r^{k}} \biggr].
\]

Following the same geometric method as before, the shadow radius seen by a distant observer at $r_{0} \to \infty$ is given by
\begin{equation}
    R_{sh} = h(r_{p}) 
    = \sqrt{ \frac{r_{p}^{2}}{f(r_{p})} \biggl[ 1 - f(r_{p}) \frac{P_{0}}{\omega_{0}^{2} r_{p}^{k}} \biggr] }.
\end{equation}
\noindent
This expression shows explicitly how the plasma parameters $P_{0}$ and $k$ modify the shadow radius through the additional radius-dependent term involving $r_{p}^{-k}$.\\

\begin{figure}[H]
	\begin{center}
		{\includegraphics[width=0.4\textwidth]{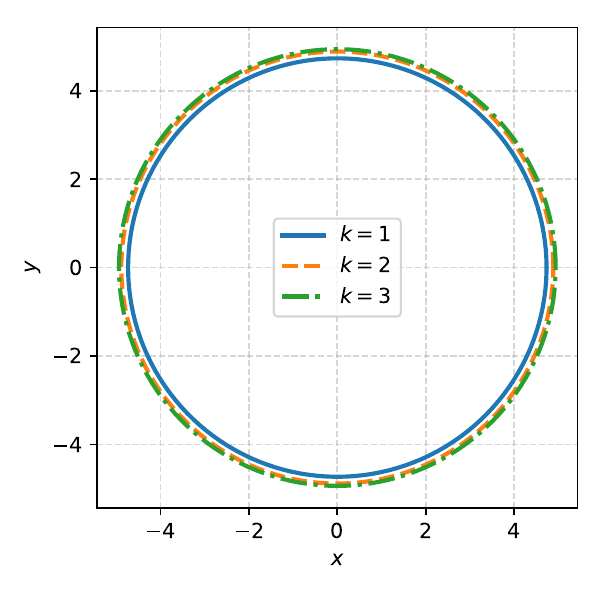}}
            {\includegraphics[width=0.4\textwidth]{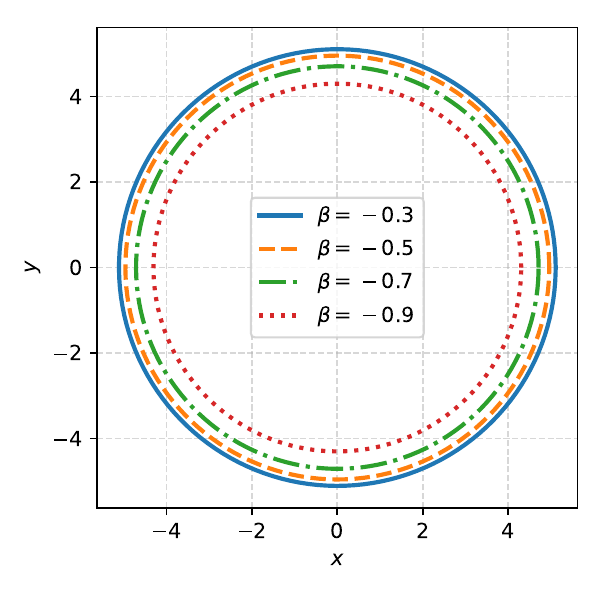}}
            {\includegraphics[width=0.4\textwidth]{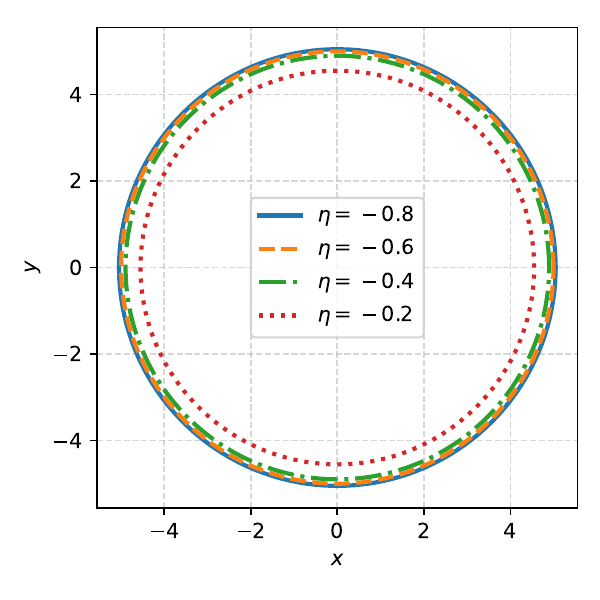}}
	\end{center}
	\caption{Variation of the shadow radius for different values of the non-homogeneous plasma parameter, $\beta$, and $\eta$ with $M=1$. 
We fix $k=2$, $\beta=-0.5$, and $\eta=-0.5$ where these parameters are not varied in the respective plots.} \label{fig:shadownonhomogeneous}
\end{figure}

The results for non-homogeneous plasma medium are illustrated in Fig.~\ref{fig:shadownonhomogeneous}. 
The overall behavior closely resembles that observed in the homogeneous plasma case. 
Notably, it is observed that the relative variation in the shadow radius with respect to changes in the plasma parameter is larger in the homogeneous plasma medium than in the non-homogeneous case. 
In addition, the analysis indicates that the shadow radius in the presence of a non-homogeneous plasma distribution is slightly larger than that for the homogeneous plasma case. 
This difference arises because the spatial distribution of plasma density in the non-homogeneous medium affects the bending of light differently, leading to a marginally larger apparent shadow.

\subsection{\textbf{Observational constraints}}
\begin{figure}[htbp]
	\begin{center}
		{\includegraphics[width=0.4\textwidth]{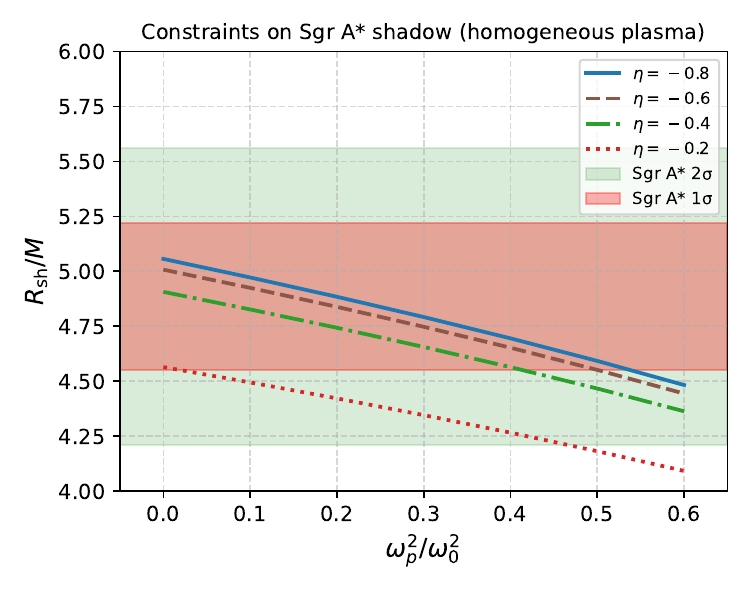}}
            {\includegraphics[width=0.4\textwidth]{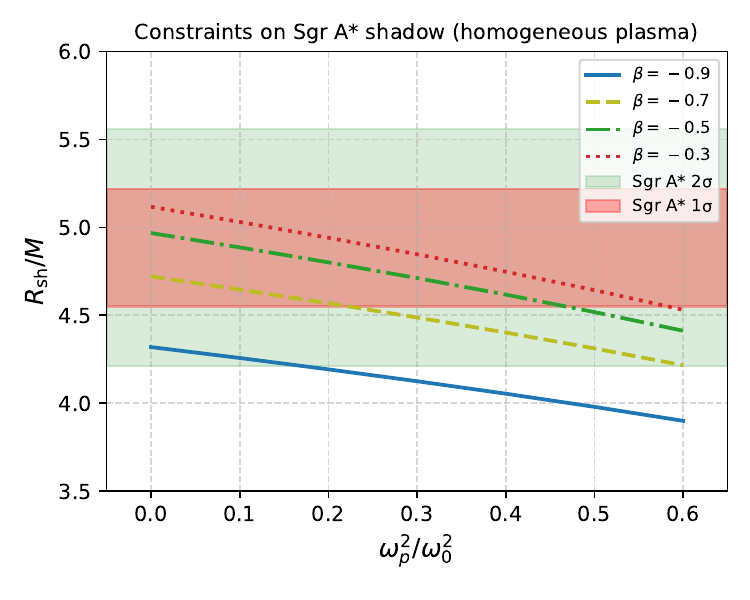}}
	\end{center}
	\caption{Shadow radius as a function of the homogeneous plasma parameter, shown for different values of the BH parameters \(\eta\) (upper panel) and \(\beta\) (lower panel). The theoretical curves are constrained by the Sgr~A* shadow radius measured by the EHT, with the shaded bands indicating the observational \(1\sigma\) and \(2\sigma\) confidence intervals.}
 \label{fig:ConstHomo1}
\end{figure}

\begin{figure}[htbp]
	\begin{center}
		{\includegraphics[width=0.4\textwidth]{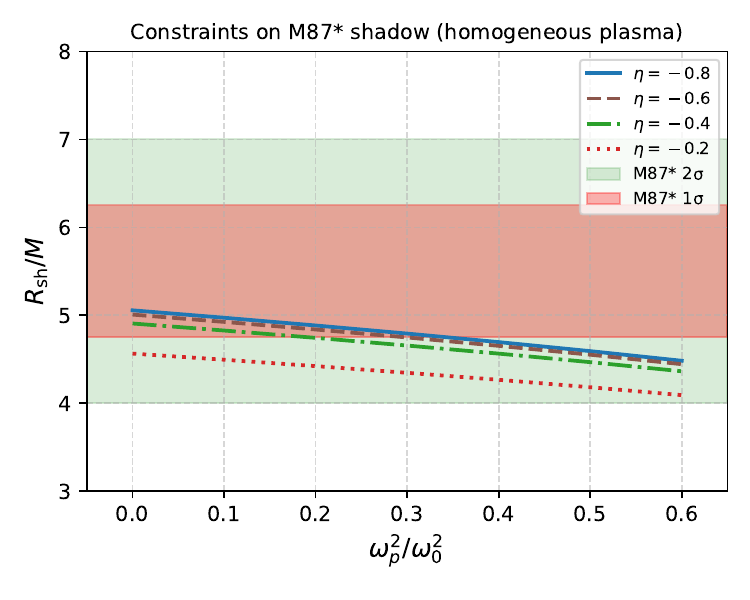}}
            {\includegraphics[width=0.4\textwidth]{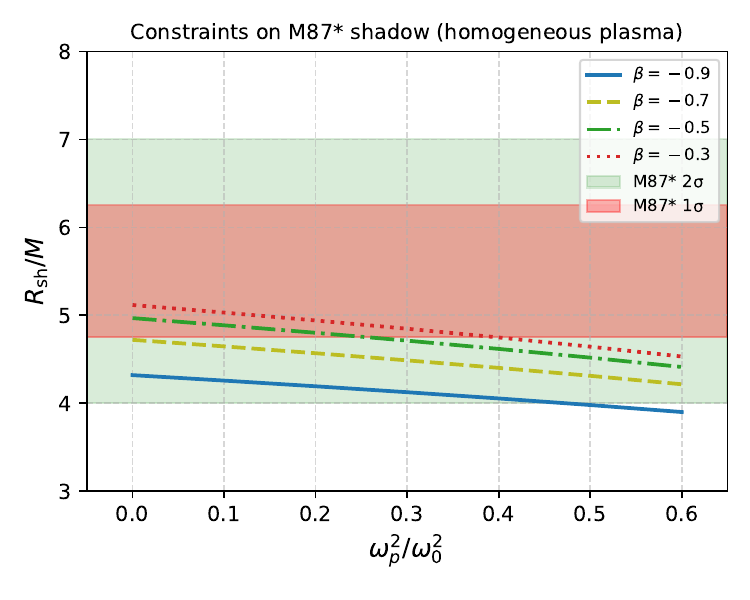}}
	\end{center}
	\caption{Shadow radius as a function of the homogeneous plasma parameter, shown for different values of the BH parameters \(\eta\) (upper panel) and \(\beta\) (lower panel). The theoretical curves are constrained by the M87* shadow radius measured by the EHT, with the shaded bands indicating the observational \(1\sigma\) and \(2\sigma\) confidence intervals.} \label{fig:ConstHomo2}
\end{figure}

\begin{figure}[htbp]
	\begin{center}
		{\includegraphics[width=0.4\textwidth]{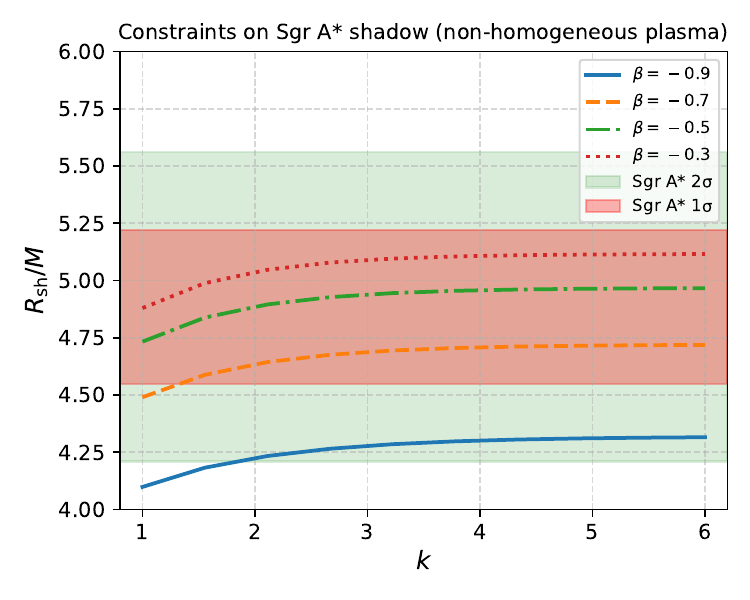}}
            {\includegraphics[width=0.4\textwidth]{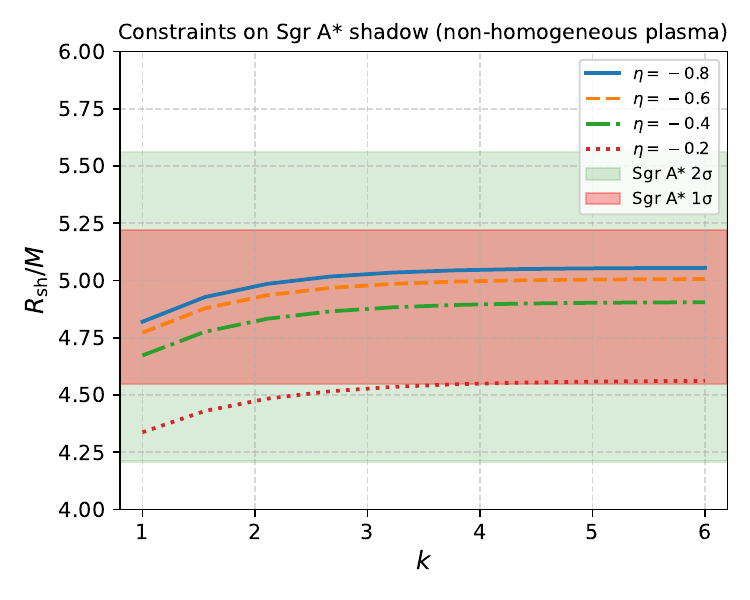}}
	\end{center}
	\caption{Shadow radius as a function of the non-homogeneous plasma parameter, shown for different values of the BH parameters \(\eta\) (upper panel) and \(\beta\) (lower panel). The theoretical curves are constrained by the Sgr~A* shadow radius measured by the EHT, with the shaded bands indicating the observational \(1\sigma\) and \(2\sigma\) confidence intervals.} \label{fig:ConstNonHomo1}
\end{figure}

\begin{figure}[htbp]
	\begin{center}
		{\includegraphics[width=0.4\textwidth]{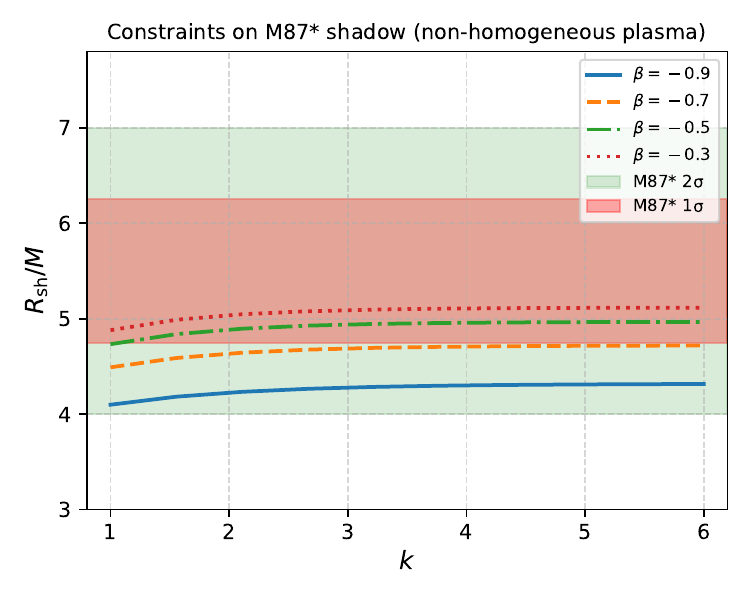}}
            {\includegraphics[width=0.4\textwidth]{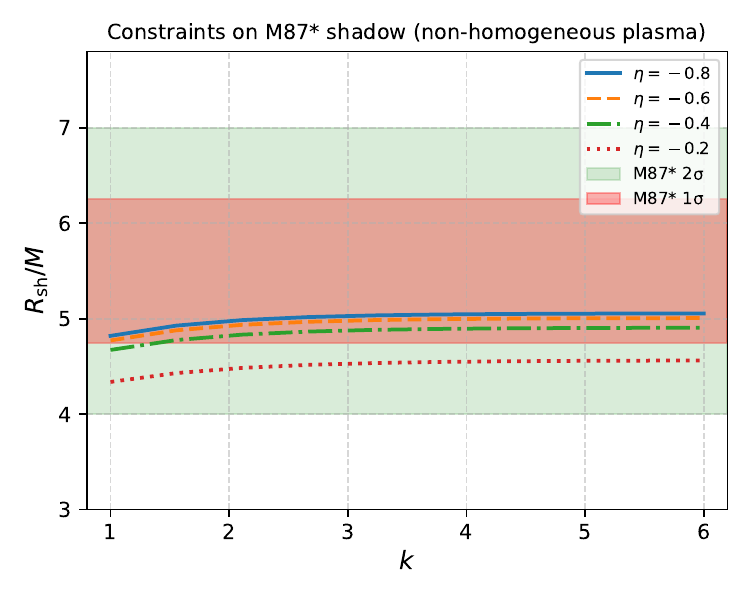}}
	\end{center}
	\caption{Shadow radius as a function of the non-homogeneous plasma parameter, shown for different values of the BH parameters \(\eta\) (upper panel) and \(\beta\) (lower panel). The theoretical curves are constrained by the M87* shadow radius measured by the EHT, with the shaded bands indicating the observational \(1\sigma\) and \(2\sigma\) confidence intervals.} \label{fig:ConstNonHomo2}
\end{figure}

In terms of observation, the BH shadow diameter plays a crucial role as it can be directly constrained through high-resolution very long baseline interferometry (VLBI) observations, most notably those performed by the EHT Collaboration \cite{Do:2019txf,GRAVITY:2020gka}. By imaging the vicinity of supermassive BHs such as Sgr~A* and M87*, the EHT has provided unprecedented measurements of the shadow size, which in turn allow us to constrain the BH parameters \( \eta \) and \( \beta \) in the presence of surrounding plasma, based on these observational results. These measurements represent one of the most direct probes of the strong gravity regime near the event horizon, and serve as an important testbed for exploring deviations from the predictions of the Kerr metric. Based on the observed angular diameters and taking into account the mass estimates of these BHs, the corresponding bounds on the dimensionless shadow radius \( r_{\text{sh}}/M \) can be deduced as follows \cite{EventHorizonTelescope:2022xqj}:
\begin{align*}
\textbf{Sgr A*}: \quad
& 
\begin{cases}
4.55 \lesssim r_{\text{sh}}/M \lesssim 5.22 & (1\sigma) \\
4.21 \lesssim r_{\text{sh}}/M \lesssim 5.56 & (2\sigma)
\end{cases}
\\[6pt]
\textbf{M87*}: \quad
& 
\begin{cases}
4.75 \lesssim r_{\text{sh}}/M \lesssim 6.25 & (1\sigma) \\
4 \lesssim r_{\text{sh}}/M \lesssim 7 & (2\sigma)
\end{cases}
\end{align*}
From Fig.~\ref{fig:ConstHomo1} to Fig.~\ref{fig:ConstNonHomo2}, we perform a detailed analysis to constrain the BH shadow by varying both the BH parameters \( \eta \) and \( \beta \) as well as the plasma parameters. 


\begin{table}[h!]
\centering
\caption{Constraints on plasma parameters ($\omega_p^2/\omega_0^2 \quad \& \quad k$) from EHT observations of Sgr A* and M87* shadows at 1$\sigma$ and 2$\sigma$ confidence levels, for homogeneous and non-homogeneous plasma media.}
\label{tab:plasma_constraints}

\renewcommand{\arraystretch}{2} 
\Large  
\resizebox{1\linewidth}{!}{   
\begin{tabular}{|c|c|c|c|}
\hline
\textbf{Case} & \textbf{1$\sigma$ bounds} & \textbf{2$\sigma$ bounds} & \textbf{Parameter ranges} \\ 
\hline
\multicolumn{4}{|c|}{\textbf{Sgr A* (Homogeneous plasma)}} \\
\hline
$\eta$ variation & $0 < \omega_p^2/\omega_0^2 < 0.4$ & $0.01 < \omega_p^2/\omega_0^2 < 0.46$ & $\eta$ in $[-0.6, -0.2]$ at 1$\sigma$; $\eta=-0.2$ at 2$\sigma$ \\ 
\hline
$\beta$ variation & $0 < \omega_p^2/\omega_0^2 < 0.2$ & $0 < \omega_p^2/\omega_0^2 < 0.15$ & $\beta$ in $[-0.7, -0.3]$ at 1$\sigma$; $\beta=-0.9$ at 2$\sigma$ \\ 
\hline
\multicolumn{4}{|c|}{\textbf{Sgr A* (Non-homogeneous plasma)}} \\
\hline
$\eta$ variation & $1 < k < 6$ & $1 < k < 3.7$ & $\eta$ in $[-0.8, -0.4]$ at 1$\sigma$; $\eta=-0.2$ at 2$\sigma$ \\ 
\hline
$\beta$ variation & $1.5 < k < 6$ & $2.4 < k < 6$ & $\beta$ in $[-0.7, -0.3]$ at 1$\sigma$; $\beta=-0.9$ at 2$\sigma$ \\  
\hline
\multicolumn{4}{|c|}{\textbf{M87* (Homogeneous plasma)}} \\
\hline
$\eta$ variation & $0 < \omega_p^2/\omega_0^2 < 0.4$ & $0.01 < \omega_p^2/\omega_0^2 < 0.46$ & $\eta$ in $[-0.6, -0.2]$ at 1$\sigma$; $\eta=-0.2$ at 2$\sigma$ \\ 
\hline
$\beta$ variation & $0 < \omega_p^2/\omega_0^2 < 0.2$ & $0 < \omega_p^2/\omega_0^2 < 0.15$ & $\beta$ in $[-0.7, -0.3]$ at 1$\sigma$; $\beta=-0.9$ at 2$\sigma$ \\ 
\hline
\multicolumn{4}{|c|}{\textbf{M87* (Non-homogeneous plasma)}} \\
\hline
$\eta$ variation & $1.7 < k < 0.4$ & $1 < k < 0.6$ & $\eta$ in $[-0.8, -0.4]$ at 1$\sigma$; $\eta=-0.2$ at 2$\sigma$ \\ 
\hline
$\beta$ variation & $1.4 < k < 6$ & $1 < k < 2.3$ & $\beta$ in $[-0.7, -0.3]$ at 1$\sigma$; $\beta=-0.9$ at 2$\sigma$ \\ 
\hline
\end{tabular}
} 

\end{table}
\vspace{0.6cm}
\noindent Table~\ref{tab:plasma_constraints} presents the constraints on the plasma parameters (\(\omega_p^2/\omega_0^2\) and \(k\)) derived from EHT observations of the Sgr~A* and M87* shadows, considering both homogeneous and non-homogeneous plasma models. The analysis shows that the BH parameters \(\eta\) and \(\beta\) can be effectively constrained within narrow ranges at the \(1\sigma\) and \(2\sigma\) confidence levels, reflecting the sensitivity of shadow observables to the surrounding plasma properties. These results highlight the potential of horiz-on-scale imaging to probe not only the BH metric but also the astrophysical environment in the vicinity of the event horizon. Overall, the study emphasizes the synergy between theoretical modeling and high-resolution observations in testing gravitational physics near supermassive BHs. Furthermore, the deviation from pure Horndeski gravity to quartic gravity in a plasma medi-um can be analyzed by comparing our results with those of Vignozzi et al. \cite{Vagnozzi:2022moj}.

\section{Shadow images with infalling gas in a plasma medium}\label{sec:8}
We consider a simplified model of an optically thin, radially infalling accretion flow around the BH in plasma medium. We adopt standard assumptions for the emission mechanism that allow us to compute the observed intensity distribution arising from the accretion flow \cite{Huang:2024bbs}. The observed specific intensity \( I_{\nu_0} \) at observer-frame photon frequency \( \nu_{\text{obs}} \), evaluated at image-plane coordinates \((X, Y)\), and typically measured in units of erg s\(^{-1}\) cm\(^{-2}\) sr\(^{-1}\) Hz\(^{-1}\), can be expressed as \cite{Jaroszynski:1997bw,Kala:2025xnb}:
\begin{equation}
    I_{\text{obs}}(\nu_{\text{obs}}, X, Y) = \int_{\gamma} g^3 \,\mathcal{J}(\nu_{\text{em}}) \, d\ell_{\text{em}},
\end{equation}
where:
\begin{itemize}
    \item \( g = \nu_{\text{obs}}/\nu_{\text{em}} \) is the redshift factor,
    \item \( \nu_{\text{em}} \) is the photon frequency in the emitter's rest frame,
    \item \( \mathcal{J}(\nu_{\text{em}}) \) is the emissivity per unit volume measured by the emitter,
    \item \( d\ell_{\text{em}} = p_\rho u_{\text{em}}^{\rho} d\lambda \) is the infinitesimal proper length element in the emitter's frame, \( p^\mu \) being the photon 4-momentum, \( u_{\text{em}}^\mu \) the emitter's 4-velocity, and \( \lambda \) the affine parameter along the photon trajectory \(\gamma\).
\end{itemize}

The redshift factor \( g \) can be determined via:
\begin{equation}
    g = \frac{p_\rho u_{\text{obs}}^{\rho}}{p_\sigma u_{\text{em}}^{\sigma}},
\end{equation}
where \( u_{\text{obs}}^\mu \) is the observer's 4-velocity.

To simplify the calculation, we restrict to equatorial photon trajectories and assume the emitting plasma is radially infalling, with 4-velocity components:
\begin{align}
u^t_{\text{em}} &= \frac{1}{g_{tt}(r)}, \nonumber \\
u^r_{\text{em}} &= -\sqrt{\frac{1-g_{tt}(r)}{ g_{tt}(r) g_{rr}(r)}}, \nonumber \\
u^{\theta}_{\text{em}} &= 0, \nonumber \\
u^{\phi}_{\text{em}} &= 0,
\end{align}
where \( g_{tt}(r) \) and \( g_{rr}(r) \) are metric functions corresponding to the BH solution under consideration.

A useful relation between components of the photon 4-momentum follows:
\begin{equation}
    \frac{p_r}{p_t} = \pm \sqrt{ g_{rr} \left( \frac{1}{g_{tt} } - \frac{\tilde{b}^2}{g_{\phi\phi}} \right) },
\end{equation}
where \( \tilde{b} \) is the photon’s impact parameter, and the sign corresponds to whether the photon is moving toward or away from the BH.

Using this, the redshift factor simplifies to:
\begin{equation}
    g = \frac{1}{ \frac{1}{g_{tt}} \pm \frac{p_r}{p_t} \sqrt{ \frac{1-g_{tt}}{g_{tt} g_{rr} } } }.
\end{equation}

For the emissivity, we assume monochromatic emission at rest-frame frequency \(\nu_*\) with a radial profile that falls as \(1/r^2\):
\begin{equation}
    \mathcal{J}(\nu_{\text{em}}) \propto \frac{\delta_D(\nu_{\text{em}} - \nu_*)}{r^2},
\end{equation}
where \(\delta_D\) is the Dirac delta function.

The proper length element along the photon trajectory becomes:
\begin{equation}
    d\ell_{\text{em}} = p_\rho u_{\text{em}}^{\rho} d\lambda = -\frac{p_t}{ g |p^r| } dr.
\end{equation}

Integrating over all observed frequencies, the total observed intensity in the image plane reduces to:
\begin{equation}
    I_{\text{obs}}(X,Y) \propto -\int_{\gamma} \frac{g^3 p_t}{r^2 p^r} dr.
\end{equation}

\begin{figure}[htbp]
	\begin{center}
		{\includegraphics[width=0.5\textwidth]{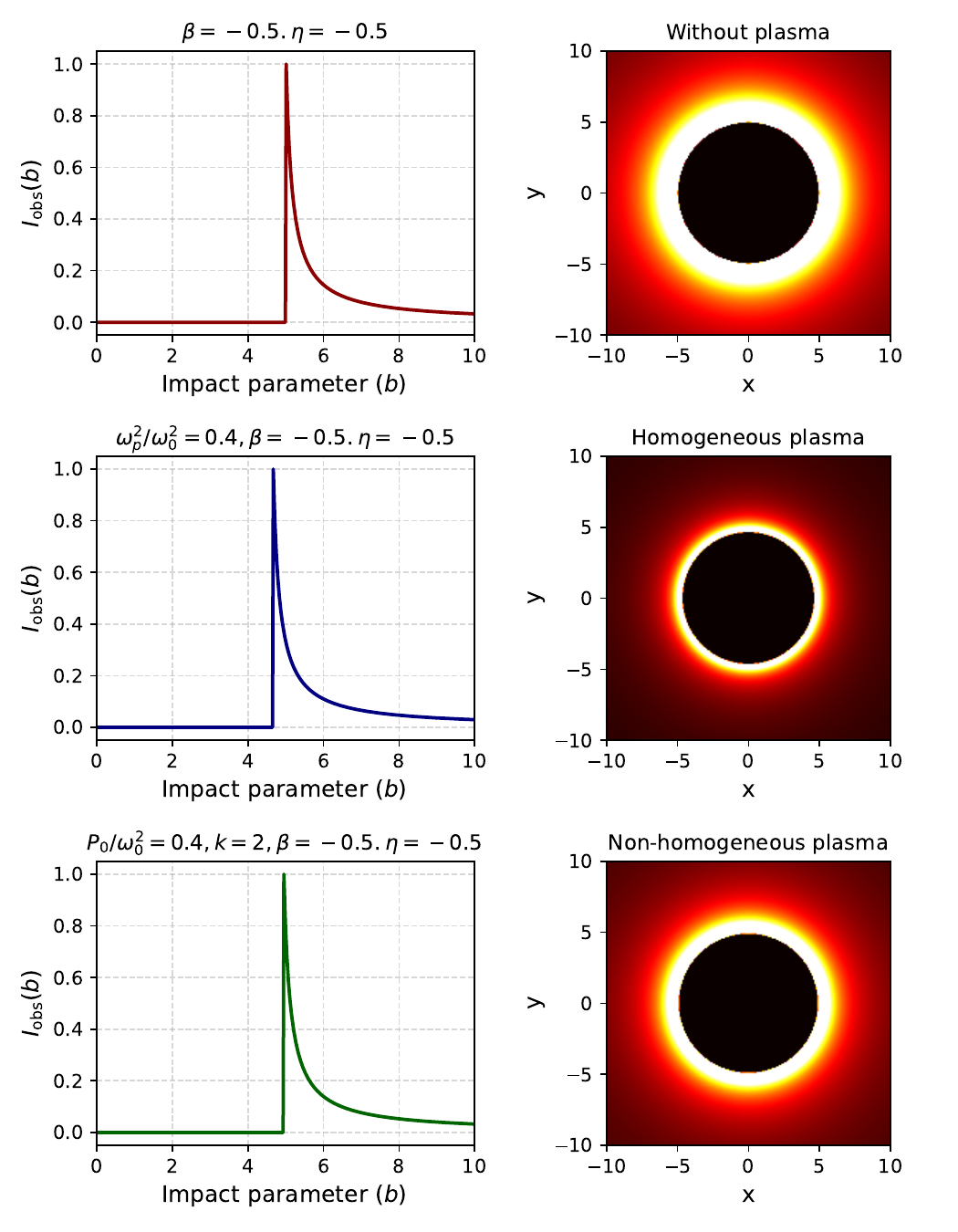}}
	\end{center}
	\caption{The left column shows the observed intensity distribution as a function of the impact parameter, while the right column presents the corresponding images of the optically thin emission region surrounding the BH. The comparison includes the cases without plasma, with homogeneous plasma, and with non-homogeneous plasma, with the parameter values indicated in the plots. X and Y represents the
angular celestial coordinates in the observer’s sky} \label{fig:shadowinfalling1}
\end{figure}

This final expression forms the basis for calculating the observed shadow and intensity map, enabling us to visualize how the plasma environment and BH parameters influence the appearance of the accreting system. We closely follow the numerical technique presented in \cite{Kala:2020prt,Saurabh:2020zqg}.
\noindent
In Fig.~\ref{fig:shadowinfalling1}, the left column displays the observed intensity distribution as a function of the impact parameter $b$, while the right column shows the corresponding shadow images of the optically thin emission region surrounding the BH. The comparison considers three scenarios: without plasma, with homogeneous plasma, and with non-homogeneous plasma, as indicated by the parameter values shown in the plots.

\begin{figure}[htbp]
	\begin{center}
		{\includegraphics[width=0.5\textwidth]{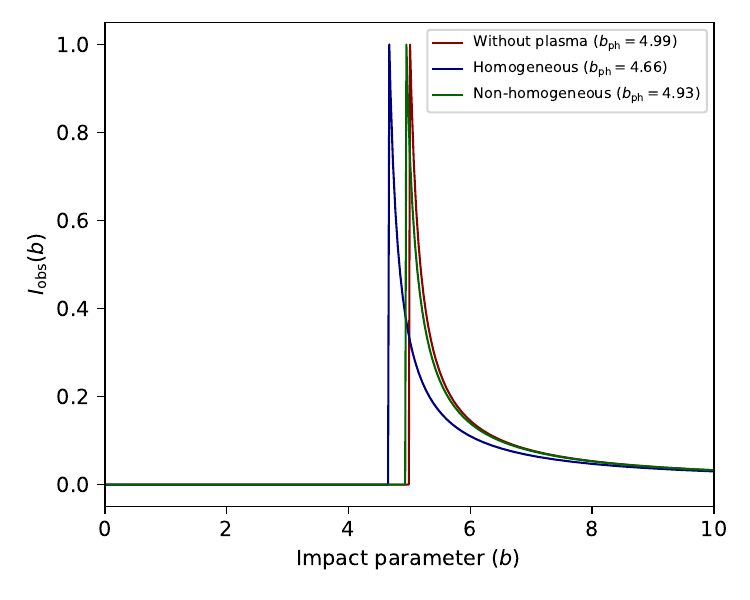}}
	\end{center}
	\caption{The observed intensity distribution as a function of the impact parameter. The comparison includes the cases without plasma, with homogeneous plasma, and with non-homogeneous plasma.} \label{fig:shadowinfalling1.1}
\end{figure}

In the combined intensity profile plot (Fig.~\ref{fig:shadowinfalling1.1}), we directly compare the flux distributions across these three cases. The vertical peak intensity is normalized for clarity; however, the location of the peak along the impact parameter $b$ exhibits a clear systematic shift reflecting the influence of plasma. For the homogeneous plasma medium, the peak appears at the lowest value of $b$, indicating that the increased refractive index pulls the photon sphere radius inward, allowing light to orbit closer to the BH. In the non-homogeneous plasma case, where the plasma density decreases with radius, the peak shifts slightly outward to a higher $b$, corresponding to a moderately larger photon sphere radius. Finally, in the vacuum case without plasma, the peak occurs at the largest $b$, consistent with the standard photon sphere determined purely by the BH geometry.

\begin{figure}[htbp]
	\begin{center}
		{\includegraphics[width=0.5\textwidth]{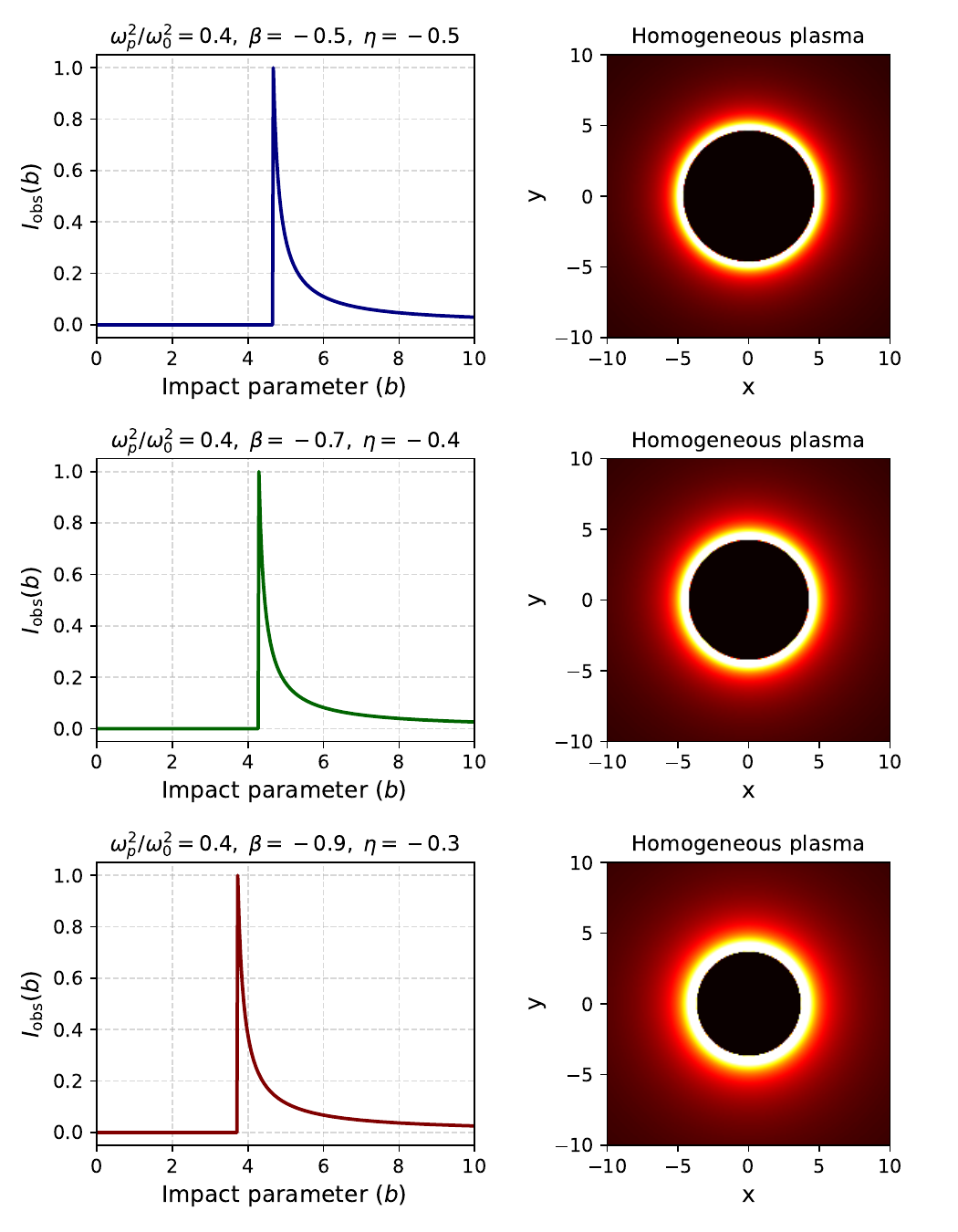}}
	\end{center}
	\caption{The left column shows the intensity distribution as a function of the impact parameter, and the right column presents the images of the BH shadow with an optically thin emission region in the presence of a homogeneous plasma medium. The plots illustrate the effect of varying $\beta$ and $\eta$, while keeping the homogeneous plasma parameter fixed, with specific parameter values indicated in the plots.} \label{fig:shadowinfalling2}
\end{figure}

\begin{figure}[htbp]
	\begin{center}
		{\includegraphics[width=0.5\textwidth]{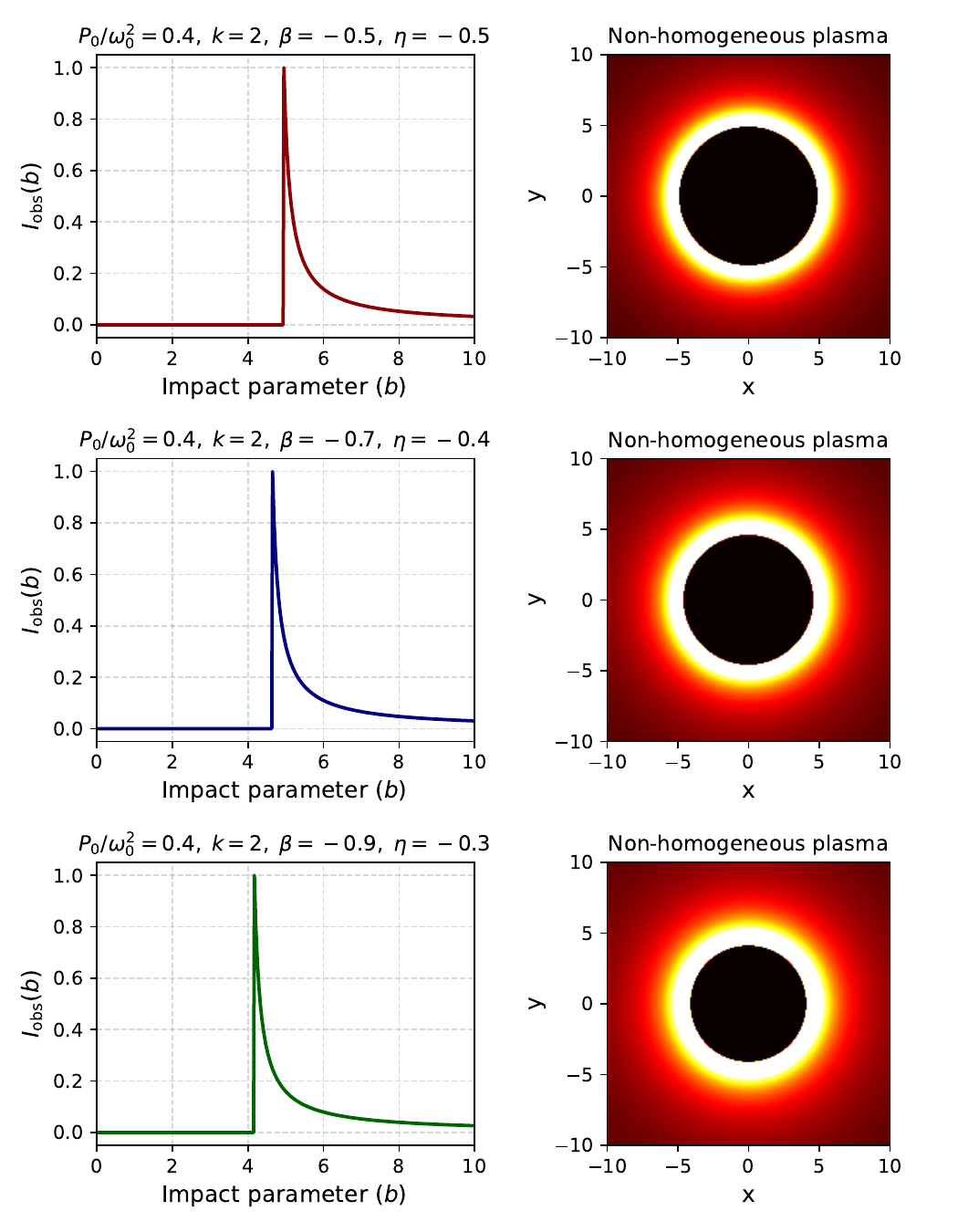}}
	\end{center}
	\caption{The left column shows the intensity distribution as a function of the impact parameter, and the right column presents the corresponding shadow images of the BH surrounded by an optically thin emission region in the presence of a non-homogeneous plasma medium. The plots illustrate the impact of varying $\beta$ and $\eta$, with the non-homogeneous plasma parameter kept fixed, as specified in the plots.} \label{fig:shadowinfalling3}
\end{figure}

A similar analysis for different sets of $\beta$ and $\eta$ values, separately shown in Fig.~\ref{fig:shadowinfalling2} and Fig.~\ref{fig:shadowinfalling3} for the homogeneous and non-homogeneous plasma cases respectively, reveals how these BH parameters further influence the shadow radius and intensity profiles. The results clearly highlight the intricate interplay between spacetime geometry, plasma properties, and the resulting observational signatures.
Physically, this behavior arises because for $b < b_{\mathrm{ph}}$, most radiation from the accretion flow is absorbed by the BH, resulting in low observed intensity. At $b = b_{\mathrm{ph}}$, photons can orbit the BH multiple times, leading to a pronounced peak in intensity due to the accumulation of light. Beyond $b_{\mathrm{ph}}$, only refracted light contributes to the observed flux, and as $b$ increases further, this contribution diminishes, causing the intensity to decrease. The observed horizontal shift of the peak directly reflects the impact of plasma on the photon sphere location and thus on the bending of light.
\section{Conclusion and future directions}\label{sec:9}
In this paper, we have investigated the deflection of light and shadow properties of a non-minimally coupled Horndeski BH in plasma medium. The key findings of our study can be summarized as follows:

\begin{itemize}
    \item It is observed that the presence of a homogeneous plasma medium increases the deflection angle of light. Furthermore, the coupling constant~$\eta$ tends to decrease the deflection angle, while the parameter~$\beta$ leads to an increase in the deflection angle in the presence of homogeneous plasma.
    \item In the presence of non-homogeneous plasma, the deflection angle increases with higher plasma parameter~$k$, while it grows with larger~$\beta$ and decreases with larger~$\eta$, highlighting how plasma distribution and BH parameters together shape light bending.
    \newline
    \item The analysis shows that the photon sphere radius increases with larger $\eta$ and decreases with larger $\beta$, reflecting their opposite effects on the effective gravitational strength. In both homogeneous and non-homogeneous plasma media, higher plasma parameters enlarge the photon sphere by raising the refractive index, enhancing light bending. Notably, in the non-homogeneous case, larger values of $k$ concentrate plasma closer to the BH, shifting the photon sphere inward due to stronger localized lensing effects.
    \item The shadow radius decreases with increasing homogeneous plasma parameter, whereas it increases with a higher non-homogeneous plasma parameter. Additionally, the shadow radius exhibits a larger relative variation with plasma parameters in the homogeneous case than in the non-homogeneous case.
    \item The constraints derived from comparing theoretical shadow radii with EHT observations of Sgr~A* and M87* show that the homogeneous plasma parameter $\omega_p^2/\omega_0^2$ typically lies in the average range of about $0$--$0.4$ at $1\sigma$ confidence level, extending up to approximately $0.46$ at $2\sigma$. For the non-homogeneous plasma medium, the density index $k$ is constrained on average within $1$--$6$ at $1\sigma$, becoming slightly narrower, around $1$--$4$, at $2\sigma$. These results highlight that shadow measurements can place meaningful limits on plasma distribution properties near both Sgr~A* and M87*, consistent across variations in the coupling constants $\eta$ and $\beta$.
    \item The analysis of shadow images with an infalling gas distribution in a plasma medium reveals that the observed intensity near the BH horizon decreases due to the presence of the refractive plasma. Notably, this reduction is more pronounced in the homogeneous plasma case, where the constant refractive index leads to stronger absorption and deflection of light rays. In contrast, for the non-homogeneous plasma, the spatially varying plasma density causes a milder effect on intensity, as the refractive influence is less concentrated near the photon sphere.

\end{itemize}
Overall, the obtained results indicate that the non--minimally coupled Horndeski gravity significantly modifies the photon sphere and shadow properties, making them sensitive to the coupling constants $\beta$ and $\eta$ as well as to the surrounding plasma distribution. This suggests that BH observations could, in principle, serve as astrophysical probes of Horndeski-type deviations from GR. It is desirable to extend this study to rotating Horndeski BHs and more realistic magnetized plasma models. These models can further constrain the observational data and provide deeper insights into optical properties of the black holes.
\section*{Acknowledgments}
The author, SK, sincerely acknowledges IMSc for providing exceptional research facilities and a conducive environment that facilitated his work as an Institute Postdoctoral Fellow. We also acknowledge the useful discussions with Prof. Sanjay Siwach during the final stage of the project.

\bibliography{main}
\bibliographystyle{unsrt}
\end{document}